\documentclass[sigconf]{acmart}%, anonymous, review]{acmart}

\setcopyright{none}
\renewcommand\footnotetextcopyrightpermission[1]{} % removes footnote with conference information in first column
\AtBeginDocument{%
  \providecommand\BibTeX{{%
    \normalfont B\kern-0.5em{\scshape i\kern-0.25em b}\kern-0.8em\TeX}}}

\usepackage{algorithm}
\usepackage{algpseudocode}
\usepackage{amsfonts,amsthm}
\usepackage{balance}
\usepackage{bm}
\usepackage{cancel}
\usepackage{colortbl}
\usepackage{graphicx}
\usepackage{hyperref}
\usepackage{subcaption}
\usepackage{xcolor}
\usepackage{xspace}
\usepackage{tikz}
\usepackage{titlesec}
\titlespacing{\paragraph}{%
  0pt}{%              left margin
  0.5\baselineskip}{% space before (vertical)
  1em}%               space after (horizontal)

\input{problem_box.sty}

\newtheorem{theorem}{Theorem}
\theoremstyle{definition}
\newtheorem{defn}[theorem]{Definition}

\begin{document}

%% The "title" command has an optional parameter,
%% allowing the author to define a "short title" to be used in page headers.
\title{Reducing Access Disparities in Networks using Edge Augmentation}
%\titlenote{This research was funded in part by the NSF under grants IIS-xxxxxxx, and IIS-xxxxxxx.}

%% The "author" command and its associated commands are used to define
%% the authors and their affiliations.
%% Of note is the shared affiliation of the first two authors, and the
%% "authornote" and "authornotemark" commands
%% used to denote shared contribution to the research.

\author{Ashkan Bashardoust}
\email{ashkanb@cs.utah.edu}
\affiliation{%
  \institution{University of Utah}
  \city{Salt Lake City}
  \state{UT}
  \country{USA}
}

\author{Sorelle A. Friedler}
\email{sorelle@cs.haverford.edu}
\affiliation{%
  \institution{Haverford College}
  \city{Haverford}
  \state{PA}
  \country{USA}
}

\author{Carlos E. Scheidegger}
\email{cscheid@cscheid.net}
\affiliation{%
  \institution{University of Arizona}
  \city{Tucson}
  \state{AZ}
  \country{USA}
}

\author{Blair D. Sullivan}
\email{sullivan@cs.utah.edu}
\affiliation{%
  \institution{University of Utah}
  \city{Salt Lake City}
  \state{UT}
  \country{USA}
}

\author{Suresh Venkatasubramanian}
\email{suresh_venkatasubramanian@brown.edu}
\affiliation{%
  \institution{Brown University}
  \city{Providence}
  \country{USA}
  \state{RI}
}

%% By default, the full list of authors will be used in the page
%% headers. Often, this list is too long, and will overlap
%% other information printed in the page headers. This command allows
%% the author to define a more concise list
%% of authors' names for this purpose.
\renewcommand{\shortauthors}{Bashardoust et al.}

\newcommand{\minadv}{broadcast}
\newcommand{\Minadv}{Broadcast}
\newcommand{\aveadv}{influence}
\newcommand{\Aveadv}{Influence}
\newcommand{\ctradv}{control}
\newcommand{\Ctradv}{Control}
\newcommand{\Ctrpair}{Pair Control}
\newcommand{\ctrpair}{pair control}

%BDS switched to terminal type for heuristic names; I like this for algorithms in general.
\newcommand{\init}{\texttt{init}}
\newcommand{\rand}{\texttt{rand}}
\newcommand{\diamcent}{\texttt{diam-both}}
\newcommand{\diampair}{\texttt{diam-chord}}
\newcommand{\mincent}{\texttt{bc-one}}
\newcommand{\mincentpair}{\texttt{bc-both}}
\newcommand{\minpair}{\texttt{bc-chord}}
\newcommand{\minave}{\texttt{infl}}

\newcommand{\probname}[1]{\textnormal{\textsc{#1}}\xspace}
\newcommand{\maxwelfare}{\probname{MaxWelfare-Augmentation}}

\newcommand{\rulesep}{\unskip\ \vrule\ }

\newcommand{\dist}{$\text{dist}^*$}
\newcommand{\diam}{$\text{diam}^*$}
\newcommand{\cent}{$\text{cent}^*$}

\definecolor{Gray}{rgb}{0.73, 0.73, 0.73}
\definecolor{Magneta}{rgb}{0.93, 0.2, 0.47}
\definecolor{Orange}{rgb}{0.93, 0.47, 0.20}
\definecolor{Teal}{rgb}{0, 0.6, 0.53}

\newcolumntype{a}{>{\columncolor{Gray!50}}p{1.1cm}}
\newcolumntype{b}{>{\columncolor{Magneta!50}}p{1.1cm}}
\newcolumntype{d}{>{\columncolor{Orange!50}}p{1.1cm}}
\newcolumntype{e}{>{\columncolor{Teal!50}}p{1.2cm}} %BDS cheating for bc-chord

\newcolumntype{f}{>{\columncolor{Gray!50}}p{0.9cm}}
\newcolumntype{g}{>{\columncolor{Magneta!50}}p{0.9cm}}
\newcolumntype{h}{>{\columncolor{Orange!50}}p{0.9cm}}

%% The abstract is a short summary of the work to be presented in the
%% article.
\begin{abstract}

In social networks, a node's position is, in and of itself, a form of \emph{social capital}. Better-positioned members not only benefit from (faster) access to diverse information, but innately have more potential influence on information spread. Structural biases often arise from network formation, and can lead to significant disparities in information access based on position. Further, processes such as link recommendation can exacerbate this inequality by relying on network structure to augment connectivity.

In this paper, we argue that one can understand and quantify this social capital through the lens of information flow in the network. In contrast to prior work, we consider the setting where all nodes may be sources of distinct information, and a node's (dis)advantage takes into account its ability to access all information available on the network, not just that from a single source.
We introduce three new measures of advantage (broadcast, influence, and control), which are quantified in terms of position in the network using \emph{access signatures} -- vectors that represent a node's ability to share information with each other node in the network. We then consider the problem of improving equity by making interventions to increase the access of the least-advantaged nodes. Since all nodes are already sources of information in our model, we argue that edge augmentation is most appropriate for mitigating bias in the network structure, and frame a budgeted intervention problem for maximizing broadcast (minimum pairwise access) over the network.

Finally, we propose heuristic strategies for selecting edge augmentations %to optimize broadcast
and empirically evaluate their performance on a corpus of real-world social networks. We demonstrate that a small number of interventions can not only significantly increase the broadcast measure of access for the least-advantaged nodes (over $5$ times more than random), but also simultaneously improve the minimum influence. Additional analysis shows that edge augmentations targeted at improving minimum pairwise access can also dramatically shrink the gap in advantage between nodes (over $82\%$) and reduce disparities between their access signatures.

\end{abstract}

%% The code below is generated by the tool at http://dl.acm.org/ccs.cfm.
%% Please copy and paste the code instead of the example below.


%% Keywords. The author(s) should pick words that accurately describe
%% the work being presented. Separate the keywords with commas.

%% This command processes the author and affiliation and title
%% information and builds the first part of the formatted document.
\maketitle

\section{Introduction}
\label{sec:introduction}

%BDS: Drawn from welfare section. Seems more suited here.
One of the promises of a highly-connected world is an impartial spread of opinions driven by free and unbiased sources of information, leading to an equitable exposure of opinion to the wide public. On the contrary, the social network platforms currently governing news diffusion, while offering many seemingly-desired features like search, personalization, and recommendation, %BDS -ing cut for line overflow
are reinforcing the centralization of information spread and the creation of so-called echo chambers and filter bubbles~\cite{Becker2020}.
A person's position within these networks often determines their access to information and opportunities such as jobs, education, and health information~\cite{colemanmedical, burt1987cohesion} and can confer advantage via influence on others~\cite{granovetter77strength}. Network position can therefore be viewed as a form of \emph{social capital}
~\cite{burt2000socialcapital, burt2004holes} -- a function of social structure that produces advantage~\cite{coleman1988social}.\looseness-1

The dynamics of how social networks are formed (including organic growth and recommendations) can lead to skews in network position based on demographics, gender, or other attributes. Experiments show that introducing even slight demographic bias to network formation processes can exacerbate differences in network structure between groups~\cite{stoicaGlass}. This becomes even more problematic when seen in light of boyd, Levy, and Marwick's argument~\cite{boyd2014networked} that position in the network is itself a \emph{feature} that can lead to discrimination separately from individual demographic attributes, and modern social networks might be vehicles for a more direct propagation of (dis)advantage. Social networks' topology can cause better-positioned users to benefit more from the privileges of their position, leading to even better connections. On the other hand, less well-connected individuals -- because of demographics, class, wealth, or other factors that drive network position -- will find it much harder to improve their network status. As a result, the gap in power between the most and least advantaged users can lead to a cascading cycle where those with more capital have better opportunities for additional improvement, creating increased inequality.\looseness-1
%BDS: I don't like this paragraph. It seems really redundant in several spots. E.g. "information access as a formal description for the access to information."

In order to mitigate the differential accumulation of social capital, one could consider intervening in the network to change the spread of information.
However, in order to do this in an automated fashion, we need ways to measure social capital based on network position. Fish et al.~\cite{Fish19Gaps} first introduced the notion of \emph{information access} as a resource and used it to propose a formal description for an individual's access to information. Beilinson et al.~\cite{beilinson2020clustering} expanded on this concept and defined an \emph{access signature} to encode the "view" from a node of its access to information sent from other nodes in the network. We build on these approaches to model structural access advantage and formulate appropriate metrics for its evaluation. We design intervention strategies that use these metrics to achieve our main goal of ensuring equitable information access.

Our setup differs from prior work in a significant way. In influence maximization, a single piece of information is being spread in the network, and one can improve access for disadvantaged nodes by augmenting the set of initial sources. In contrast, we consider a setting such as those which occur on LinkedIn, where each node is the source of a unique piece of information, and access to all pieces is equally important. Given this
key difference, we argue that instead of trying to select additional seeds for some or all of the pieces to improve dispersal, the natural intervention is adding edges to the network, representing the idea of purposefully strengthening weak ties~\cite{granovetter77strength} to mitigate bias in the structure and increase connectivity.\looseness-1

In this work, we have three primary contributions:

(1) Using a normative framework and drawing on prior work, we formulate three measures -- \emph{\minadv{}}, \emph{\aveadv{}}, and \emph{\ctradv{}} -- to model structural advantage with respect to access.

(2) We focus on intervening in the network using budgeted edge augmentation to improve the structural position of least-advantaged nodes, reduce the advantage gap, and ensure that nodes have similar ``views'' of the network (as measured via their access signature). At the core of our approach is the idea that
to mitigate inequality, we should maximize the minimum access of the least-advantaged node -- which in turn reduces to maximizing the minimum access between all pairs of nodes in the network.

(3) We introduce heuristic algorithms for selecting edge augmentations and empirically evaluate them on a corpus of social network data. We further show experimentally that while this process directly maximizes the \emph{\minadv{}} measure of access advantage, it also simultaneously improves \emph{\aveadv{}} and \emph{\ctradv{}} disparities among nodes, as well as making node access signatures more uniform.
%BDS: I agree that this is weak as written and needs to be improved.

\section{Related Work and Preliminaries}
\label{sec:prerelated}

%We begin by surveying related work on influence maximization and
%the impact of network structure on information flow, including evidence for structural bias and prior investigations of algorithmic fairness when information is being propagated on a network.

Motivated by the design of viral marketing strategies, Domingos and Richardson~\cite{domingos01mining} introduced an algorithmic problem for social networks in which one wished to convince an initial subset of individuals to adopt a new product or innovation in order to maximize the cascade of further adoptions. This model can be generalized to many types of information spread beyond adoption and was formalized as the discrete optimization problem of \emph{influence maximization} by Kempe et al.~\cite{kempe03maximizing}, leading to an extensive literature on the subject (see the survey~\citep{li2018influence}), including many applications in public health awareness
~\cite{endtoendWilder2018, bridgingYadav2018, identifyingValente2007, clinicalWiler2020}.

\paragraph{Structural Advantage}
Information propagation in networks has been studied for decades in social and computing sciences~\cite{colemanmedical, burt1987cohesion}, and network position is known to dramatically impact a node's access to other network members~\cite{granovetter77strength}. It has been
repeatedly argued that one's position in a network is itself a form of wealth or social capital~\cite{coleman1988social,burt2000socialcapital, burt2004holes, jackson19human},
enabling better and faster access to circulating information and important individuals.
This translates into better access to opportunities (such as jobs and scholarships) and enables
well-positioned people to be more effective brokers, make better decisions, and innovate more efficiently~\cite{burt2004holes}. Further, in public health scenarios, people rarely act on mass-media information unless it is also transmitted through personal ties~\cite{katz1966personal, rogers1962diffusion}, leading to well-connected nodes having improved outcomes in crises.

\paragraph{Bias in Network Structure}
The network itself can act as a transmitter for bias when the structural advantages described above interact with network formation mechanisms that encourage homophily and clustering of demographic groups. Schelling demonstrated how local neighborhood-based decisions could lead to segregation~\cite{schelling1971}, and recent work has explored how bias in localized decisions about new connections can result in networks that have significant skew~\cite{karimi2018, leskovec2008}. Sociologists have extensively studied the role of social status in shaping network structure, showing in small-scale experiments that it significantly influences whether individuals end up in central vs. peripheral network positions~\cite{campbell1986,  lin1999}.

More recently, studies in network science have extended these ideas to large-scale networks by developing computational methods for characterizing the structural influence of social status at scale~\cite{ball2012, leskovec2010}. For example, Clauset et al. quantify the ways in which institutional reputation (and the auxiliary features of demographics and productivity) shapes the structure of faculty hiring networks among academic departments~\cite{clauset2015, way2016} and subsequently the differential spread of ideas~\cite{morgan2018}.

\paragraph{Algorithmic Fairness in Information Propagation}
In the setting of information access, natural questions of fairness arise in the problem of ensuring
similar allocation among demographic groups, which are often represented as disjoint subsets of nodes.
Inspired by the literature on social position initiated by Granovetter's strength of weak ties \citep{granovetter77strength} and framed in the context of online social networks by boyd, Levy, and Marwick \citep{boyd2014networked}, there has been a rash of recent work on computational questions around fairness in access on social networks~\cite{Fish19Gaps, tsang2019group, ali2019fairness, stoica2019fairness, rahmattalabi2020, jalali2020unfairness, becker2021, Wang2021Information}. The key underlying idea is that
\emph{information access is a resource}, and Fish et al.~\cite{Fish19Gaps} argued that access based
on network position is a form of privilege, which they used to define a notion of individual fairness.
%BDS I don't understand the point of some of these related work paragraphs.

Much of the work on defining and applying fairness has been undertaken in the influence maximization framework. One important thrust has been improving equity among demographic groups within a network, typically defined based on protected classes (e.g., race, gender)~\cite{stoica2019fairness, tsang2019group, ali2019fairness, rahmattalabi2020, jalali2020unfairness}. They develop metrics and algorithms to ensure that roughly equal amounts of information reach each demographic group while optimizing influence maximization. %BDS I don't know what this means so I couldn't rewrite it.
In all cases, a single piece of information is being spread in the network, and they intervene by augmenting the seed set. The one exception is Jalali et al.~\cite{jalali2020unfairness} who adds edges instead of seeds. We note that while a few papers have considered edge augmentation to maximize the influence of a given group~\cite{dangelo2019recom, becker2021}, they
inherently define advantage to be access to the seed set.
%BDS I'm not happy with this, but I don't know how to improve it.

Several other recent papers in the space consider variants of the basic access problem. Becker et al.~\cite{Becker2020} consider $\mu$ sources of diverse information in a network and maximize the expected number of nodes receiving at least $v$ types of information. Ramachandran et al.\cite{Gage2020} use a diffusion model of mobility dynamics and try to achieve equity in group-level access in the facility location problem.

\subsection{Preliminaries}

As in the discrete optimization setting of~\cite{kempe03maximizing}, we use a stochastic \emph{information flow model} describing how information might transmit from one node to another along the edges of $G$ (for example, Independent Cascade, Linear Threshold, or an infection flow model from epidemiology~\cite{kempe03maximizing}). These models all work by assuming that at time zero, an initial \emph{seed set} of nodes that possess the information to be spread. For each seed $v_j$ in the seed set, there is then a (potentially hard to compute) probability $p_{ij}$ -- which we call \emph{access distance} --  that node $v_i \in V$ possesses $v_j$'s information once the spread process has terminated. Inversely, $p_{ji}$ is called the
\emph{reach} of $v_i$ with respect to $v_j$. Since we restrict our attention to the undirected setting (as social network links require mutual consent and typically create a giant connected component -- Facebook's has 99.9\% of users~\cite{Ugander2011TheAO}), $p_{ij} = p_{ji}$ and we use them interchangeably.

\paragraph{Independent Cascade Model}
In this work, we utilize the standard probabilistic model of influence propagation, Independent Cascade (IC)~\cite{kempe03maximizing} with a uniform transmission probability $\alpha$. In this model, a node exists in one of the three states: \emph{ready to receive}, \emph{ready to transmit}, or \emph{dormant}. Initially (at time zero), all nodes are ready to receive information, while the \emph{seed} nodes also possess the information and are ready to transmit. At each time step, a node that is ready to transmit sends its information to neighbors by transmitting along each incident edge independently with probability $\alpha$. All such transmissions are imagined to happen simultaneously, after which the transmitting node goes dormant. Computing the access probabilities for Independent Cascade is $\#P$-hard~\cite{sharpphard10chen}, so we use standard \emph{Monte Carlo} simulations to estimate them when needed.

\paragraph{Access signatures}
Since we view a piece of information as being uniquely identified by its originator, describing the access of a node requires a vector of $n-1$ probabilities, which is standardized to length $n$ to facilitate easy indexing and comparison across nodes, and $p_{ii} := 1$. These vectors are called \emph{information access signatures}, and were introduced by Beilinson et al.~\cite{beilinson2020clustering}, who argued that nodes that have similar ``status'' based on network position receive similar information.
%proposed the \emph{information access signature} as a way to encode the ``view'' from a node of its access to information sent from the other nodes in $G$.
The signature encodes the ``view'' from a node of its access to information sent from the other nodes in $G$;
%characterizes a node's information access based on how likely they are to receive information from everyone else in the network;
people who are likely to receive information from the same part(s) of the network will have similar signatures.

\begin{defn}[Access Signature~\cite{beilinson2020clustering}]
The \emph{access signature} $a^G_{\alpha} \colon V \to \mathbb{R}^n$ of a node $v_i \in V$ in graph $G$ on $n$ nodes is:
\[ s^G_{\alpha}(i) = (p_{i1}, ..., p_{ij}, ... p_{in}) \]
\end{defn}

\section{Structural Advantage}
\label{sec:advantage}
How does network position impact access and influence? In social networks, structural advantage can manifest in many ways. Inspired by prior work, we formalize three distinct notions of advantage arising from network position and propose measures for quantifying each.\looseness-1

\subsection{Access-based Definitions}
\label{sec:definitions}
We begin by defining analogues of graph-theoretic distance, diameter, and betweenness centrality, highlighting when the access-based variants diverge from their traditional counterparts.

\paragraph{Access Distance}
In graph theory, the \emph{distance} between nodes $v_i$ and $v_j$ is the number of edges in a shortest $v_iv_j$-path. To adapt this to an information flow setting, we let the \emph{access distance} be
\[ \text{\dist{}} (v_i, v_j) = p_{ij}, \]
the probability that $v_i$ receives $v_j$'s information after the completion of Independent Cascade.  We observe that these measures can diverge in even simple networks. Consider two nodes connected by an edge; they have distance 1 and access distance $\alpha$. If instead, these nodes were connected
by $t$ disjoint paths of length $2$ they woud have distance $2$, but access distance $1 - (1 - \alpha^{2})^t$. Assume $t > \frac{\log(1 - \alpha)}{\log(1 - \alpha^2)}$. While the nodes are graph-theoretically closer in the first scenario, in the information access setting they are closer in the second.

\paragraph{Access Diameter}
For large networks, we often rely on summary statistics as indicators of network structure. One such
metric is the \emph{diameter}, defined to be the maximum distance between any two nodes (equivalently, the length of a longest shortest path). The analogous notion in the information access setting is then
then the smallest access distance between two nodes (equivalently, the lowest probability of pairwise information transmission). We call this the \emph{access diameter}:
\[ \text{\diam{}}_{G} = \min_{v_i, v_j \in V} p_{ij}. \]

\paragraph{Access Centrality}
Finally, since we are interested in assessing influence or control with respect to information flow,
we consider the \emph{betweenness centrality}, which measures how often a node appears on the shortest paths between others. Specifically, if we let $\sigma_{jk}$ be the number of shortest $v_jv_k$-paths, and
$\sigma_{jk}(i)$ the number of shortest $v_jv_k$-paths passing through vertex $v_i$, we can
define the betweenness centrality of $v_i$ as
\[ g(v_i) = \sum_{v_j \neq v_k \neq v_i \in V} \frac{\sigma_{jk}(i)}{\sigma_{jk}}. \]

One can think of this as measuring the brokerage ability of a node in a world where information flows along the shortest paths. To adapt to the Independent Cascade model, we want to measure the fraction of other nodes' pairwise access that depends on $v_i$. In other words, the \emph{access centrality} $v_i$ is
\[\text{\cent{}} (v_i) = \sum_{v_i \neq v_j \neq v_k \in V} \frac{p_{jk} (i)}{p_{jk}}, \]
where $p_{jk}(i) = p_{jk} - p'_{jk}$ can be computed using the access distance $p'$ in $G' = G \setminus v_i$. We note this is computationally expensive, as you must re-estimate access distances in $G \setminus v$ for each vertex $v$.

To see where these two notions diverge, consider nodes $a,b$ connected with a path of length two through node $c$. The betweenness and access centrality of $c$ are both $1$. Now augment this graph by adding $t$ disjoint $ab$-paths of length $3$; the betweenness centrality of $c$ remains 1, but the access centrality tends to $0$ as $t$ increases, as the fraction of information passing through $c$ becomes insignificant.

%BDS Moved from control advantage discussion. Ashkan proposes to condense to a sentence and merge with above. Postponed.
%between the leaves, and has complete control over them (value 1, they cannot communicate without his participation). On the other hand, removing a leaf does not impact communication between the other two nodes at all, so they have control value 0.

\subsection{Measures of Advantage}
\label{sec:measures}

We now formalize three different notions of structural advantage, arising from various perspectives on fairness and information flow.

\subsubsection{\Minadv{} Advantage:}
From a fairness point of view, Fish et al.\cite{Fish19Gaps} argued that the performance of a source should be measured by how effectively it reaches least-advantaged nodes. In this vein, we propose our first advantage function, \emph{\minadv{}}, to measure how difficult it is for a node to disseminate its information to \emph{all} others in the network.

\begin{defn}[\Minadv{} Advantage]
  The \emph{\minadv{} advantage} of a node is the worst-case probability that its information is received -- equivalently, the minimum entry in its access signature:
  \[\text{\minadv{}}(v_i) = \min_{p_{ij} \in s(i)} p_{ij}. \]
\end{defn}

In some sense, this represents how ``loud'' the node is -- a larger \minadv{} means a better probability that everyone else in the network will receive your information. Consider the case of recruiters using a network like LinkedIn, wanting to spread information about a job opportunity. In order to ensure a diverse candidate pool and broad reach, the employer wants a high probability the ad will reach all suitable nodes in the network. Since well-connected users receive many such ads, the measure of recruiting effectiveness will depend on how well they can disseminate the information to the least-advantaged members of the network. Better-positioned recruiters will have higher \minadv{}.

Further, social media is often used in public health \emph{epidemiologic monitoring and surveillance} for early detection of disease outbreaks. Staff responsible for dispelling misinformation and identifying high-risk or affected groups need access not only to the majority of people, but especially to those who are poorly-connected (and thus at risk of being neglected in treatment~\cite{CDCmanual, socialmediapublichealth}), motivating us to improve their \minadv{}.

From another perspective, the \minadv{} is a lower bound on the probability that $v_i$ will get information from $v_j$, \emph{regardless of which $v_j$ is selected}! Increasing \minadv{}($v_i$) necessarily improves information flow to/from the parts of the network that are currently least accessible from $v_i$, increasing the novelty and diversity of its information. Novel information often represents a resource or opportunity due to local scarcity, and users with access to it enjoy social and economic advantages, including more success in wages, promotion, job placement, and creativity~\cite{granovetter77strength, burt2004holes}.

%BDS I don't understand how this is relevant *at all* to the definition at hand.
%Social media can be used following natural or man-made disasters to increase \emph{situational awareness} of humanitarian crises. Individuals in distress can use social media to seek help (e.g., water supply and shelter) by self-reporting their humanitarian needs and connecting with family, friends, and emergency responders. The authorities can use social media to identify individuals in distress and respond accordingly. Nongovernmental organizations can also use social media to track and map the needs of displaced people~\cite{CDCmanual, socialmediapublichealth}. Anderson et al. show that during Hurricane Sandy in 2012, most affected areas were underrepresented in the social media conversation about Sandy. They could not make their voices be heard due to their invisibility on social media, resulting in them not getting the support they needed~\cite{anderson2016farfar}.

\subsubsection{\Aveadv{} Advantage:}

Network prominence has been studied as a type of advantage~\cite{brass1992power, prominence}. A central or well-connected node is more likely to have high visibility, which Jackson's \emph{friendship paradox}
argues can lead to over-representation and increased influence~\cite{jackson19human}. This type of advantage does not require the ability to reach all nodes in the network, just many of them.

Being able to disseminate information to a large set of other members enables a user to build their social reputation, express and diffuse their opinion, and discover novel content and information~\cite{dangelo2019recom}, which can be viewed as \emph{media power} or celebrity capital. This may also lead to opportunities for revenue from advertisement~\cite{burt2013advantage}. Consider the example of collaborations in a scientific community. If someone can reach more people to share her research, she gets more recognition, and feedback which enables improvement, collaboration opportunities, and directions or ideas for future work~\cite{boyd2021knit, sugimoto2021science}.
We propose \emph{\aveadv{}} advantage as a measure of this form of structural advantage, drawing on influence maximization~\cite{kempe03maximizing} in choosing a quantification.

\begin{defn}[\Aveadv{} Advantage]  The \emph{\aveadv{} advantage} of a node is the average probability that its information is received -- equivalently, the mean of the entries in its access signature:
\[\text{\aveadv{}}(v_i) = \frac{1}{n} \sum_{p_{ij} \in s(i)} p_{ij} = \frac{1}{n} \sum_{v_j \in V} \text{\dist{}}(v_i, v_j)\]
\end{defn}

\subsubsection{\Ctradv{} Advantage:}
Burt~\cite{burt2004holes} introduced the idea of brokerage advantage. Individuals in networks with many ``structural holes'' may derive information and control benefits from the lack of external connectivity among people they can reach. Burt introduced this form of social capital as an information benefit or vision advantage that improves performance by providing early access to diverse and novel perspectives, ideas, and information. Hence, a person's reach is a form of power as it enables her to broker favors and consolidate strength by being uniquely positioned to coordinate the actions of others. We call this type of structural advantage \emph{\ctradv{}}.

While Burt proposed several ways to measure structural holes, including bridge count~\cite{Burt2000BridgeCount}, and network constraint/redundancy~\cite{burt2004holes}, in more
recent work Jackson~\cite{jackson19human} used betweenness centrality~\cite{FreemanBetween} to measure brokerage advantage. This generic measure of importance in a network captures a node's ability to act as an intermediary to coordinate others, where nodes rely on it in order to reach other users along shortest paths. Higher centrality corresponds to more control over information flow in the network. In turn, we use access centrality to measure the \ctradv{} advantage.

\begin{defn}[\Ctradv{} Advantage]
The \emph{\ctradv{} advantage} of a node is given by its access centrality:
\[\text{\ctradv{}}(v_i) = \text{\cent{}} (v_i) = \sum_{v_i \neq v_j \neq v_k \in V} \frac{p_{jk}(i)}{p_{jk}} \]
\end{defn}

We observe that \ctradv{} can be rewritten as a nested sum over nodes, revealing
a useful finer-grained notion of advantage. For example, suppose the node $v_i$ has
one neighbor $v_j$, which is a leaf, and another neighbor which is a member of a large clique.
Clearly, $v_i$ has a large degree of control over $v_j$, as it is an intermediary to all
access to the clique, yet \ctradv{}($v_i$) might remain small, as $v_i$ plays little role
in access between clique nodes. We use \ctradv{}$^{i}_{j}$ to denote the brokerage $v_i$ has over information reaching node $v_j$, where
\[\text{\ctradv{}}^{j}_{i} = \sum_{v_j \in V} \frac{p_{ij} (c)}{p_{ij}}. \]
Our measure can then be written as $\text{\ctradv{}}(v_i) = \sum_{j} \text{\ctradv}^{i}_{j}$.

When trying to mitigate inequity in access, we would like to see the \ctradv{} values decrease for
better-positioned nodes. Additionally, we argue that in an ideal network, no node has a monopoly over others' access to information, and we would like to prevent situations where \ctradv{}$^{i}_{j}$ is close to $1$ for any pair $(i,j)$.

\section{Edge Intervention \& Welfare}
\label{sec:edge-welfare}

%BDS merging sections; edge argument needs to get much shorter.
In contrast to the standard framework of influence maximization, we argue that when considering
information flow in a network, it is important to have access to information from all individuals,
not just a seed set. Further, given this shift in objective, adjustments to the model of intervention
are warranted, and we propose edge augmentation as the natural candidate. We support our argument from three perspectives: \emph{variety}, \emph{structure}, and \emph{voice}.

\paragraph{Variety} Since ideas travel a variety of paths from many sources~\cite{von1988sources, geroski2002learning, menon2003valuing}, access to more diverse information and a greater number of individuals is important~\cite{granovetter77strength} and can provide a vision advantage that translates into social capital~\cite{burt2004holes}. Key functionalities of social networks like LinkedIn rely on the fact that important information is frequently being disseminated from a multitude of constantly-changing sources. Traditional influence maximization is insufficient for assessing
access and proposing equity-improving interventions in this setting, as we no longer know the seed set, nor can we afford to try and augment sources for each new announcement.

\paragraph{Structure} Granovetter introduced the idea of network manipulation to achieve specific goals~\cite{granovetter77strength}. Since network position is a critical form of social capital in
information access, and positional disparities arise from biases in the network structure, we
argue that interventions which change the underlying connectivity of the network are necessary.
The natural candidate is to increase access through edge augmentation. This approach is further
supported when one thinks of these edges as representing the addition of weak ties to the transmission network, as research shows that information can traverse greater social distance and reach more people when diffused along weak ties instead of strong ones~\cite{granovetter77strength}.

\paragraph{Voice} While it is easy to focus on improving access for poorly-positioned nodes, it is also important to consider the effect of interventions on already-advantaged users. Specifically, node
interventions increase the reach (and thus influence) of selected individuals~\cite{jackson19human}, essentially amplifying their information within the network. To give voice to all participants, we argue that edge augmentation improves fairness by increasing the reach of all nodes.\par

\paragraph{\nopunct} Now that we have argued for using edge augmentation to intervene in the network, we turn to the question of which structural measure of advantage to optimize. We use a normative framework to select one of broadcast, influence, and control, and draw on the Rawlsian Maximin argument~\cite{rawls2009theory} in proposing that we should maximize the advantage of the least advantaged node(s).

To choose a notion of advantage, we begin by observing that optimizing \aveadv{} encourages the formation of edges to well-positioned nodes. Therefore, nodes with better connections become more attractive to connect to~\cite{jackson19human}, leading to a rich-get-richer phenomenon and potentially increasing the advantage gap instead of equalizing access~\cite{burt1999opinion}. These peripheral-central connections also increase the \ctradv{} of central nodes over others, especially the disadvantaged.
On the other hand, using \minadv{} as the objective prioritizes connectivity for the most disadvantaged
nodes. As John Stuarts Mills noted, "it is hardly possible to overrate the value . . . of placing human beings in contact with persons dissimilar to themselves and with modes of thought and action unlike those with which are familiar . . . Such communication has always been and is peculiarly in the present age, one of the primary sources of progress"~\cite{miller1848}. Optimizing for \ctradv{}, on the other hand, prioritizes the brokerage ability of nodes over their access to diverse information, which could lead to polarization and centralized information distribution. We argue that increasing \minadv{}, which tends to also reduce the \ctradv{} of other nodes, is preferable since depending on powerful information-brokers reduces one's chance of unbiased access to diverse opinions.

Several other normative reasons underlie our preference for \minadv{} to measure structural advantage, when one considers outcomes in a network containing several (mostly-disjoint) minority groups. First,
while these groups may have common interests, they will not individually have enough influence to accomplish them. Connecting disadvantaged nodes directly (instead of through a central node) will enable
them to support one another and access important information, while countering the ever-increasing power of the majority. In support of this argument, we note that Kogan et al. show that geographically vulnerable (disadvantaged) users propagate more information during disasters, and are more likely to propagate tweets from other geographically vulnerable users~\cite{Kogan2015Think}. A final argument
arises from work on mitigating polarization in social networks by increasing the similarity of
users' exposure to a broad diversity of news and ideas. Since minimizing diameter can speed up communication~\cite{demaine2010diam} and increase the uniformity of exposure times, we argue that
optimizing \minadv{} is the natural analogue in the information access setting.

%BDS I don't see where we argue for broadcast over control as an objective.
To formalize a discrete optimization problem, we must now transform our advantage measure into an objective function. Following the Rawlsian \emph{Maximin Principle} that one should maximize the welfare of the worst-off person~\cite{rawls2009theory}, we seek to maximize \minadv{} for the least-advantaged nodes, and
% Equivalently, we seek to minimize the maximum access distance between any pair of nodes in the network,
formalize this as the \emph{welfare}.

\begin{defn}
The \emph{welfare} of a graph $G=(V,E)$ is
\[\mu(G) = \min_{v_i \in V} \text{\minadv{}}(v_i) = \min_{v_i , v_j \in V} \text{\dist{}}(v_i, v_j)\]
 \end{defn}

Our central problem is to find a \emph{budgeted intervention} optimizing welfare.\looseness-1

 \begin{problembox}{MaxWelfare-Augmentation}
 	\Input & A graph $G = (V,E)$ and an integer $k \in \mathbb{N}$. \\
 	\Prob & Find a set $E^+ \subseteq V\times V$ of size at most $k$ so that
	$\mu(H)$ is maximized, where $H := (V, E \cup E^+)$.
\end{problembox}

\section{Heuristics}
\label{sec:heuristics}

%BDS Figure placement is fragile!
%Charts should support the story: overall message of the trend in the data - driven by the data
\begin{figure*}
	\centering
	\begin{subfigure}[t]{\textwidth}
  		\includegraphics[width=.24\columnwidth]{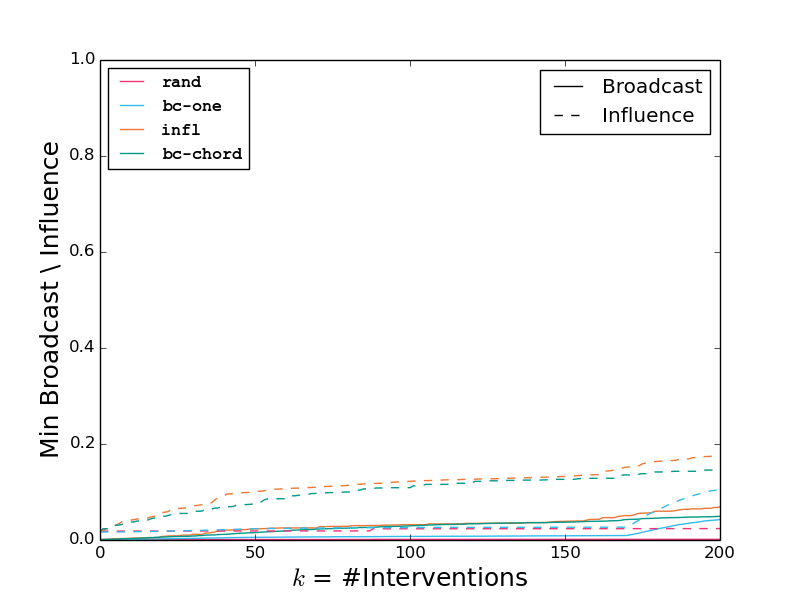}
		\includegraphics[width=.24\columnwidth]{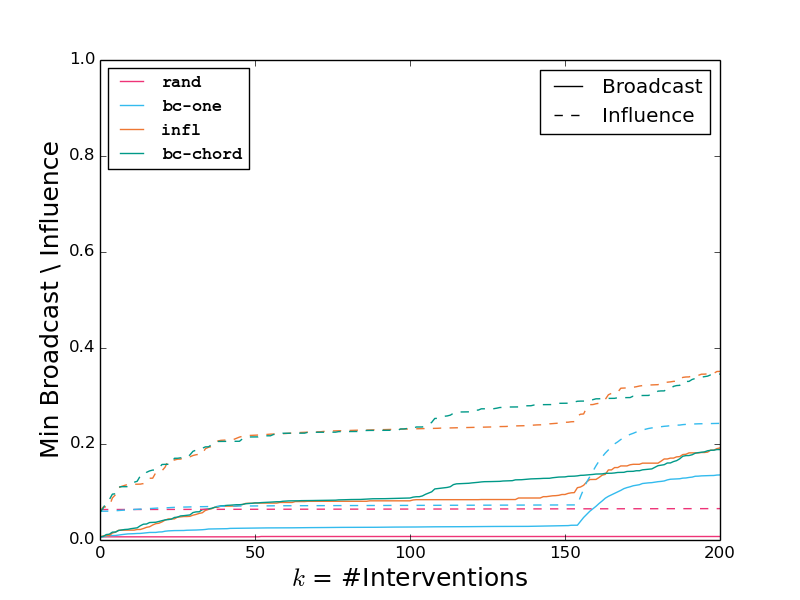}
		\includegraphics[width=.24\columnwidth]{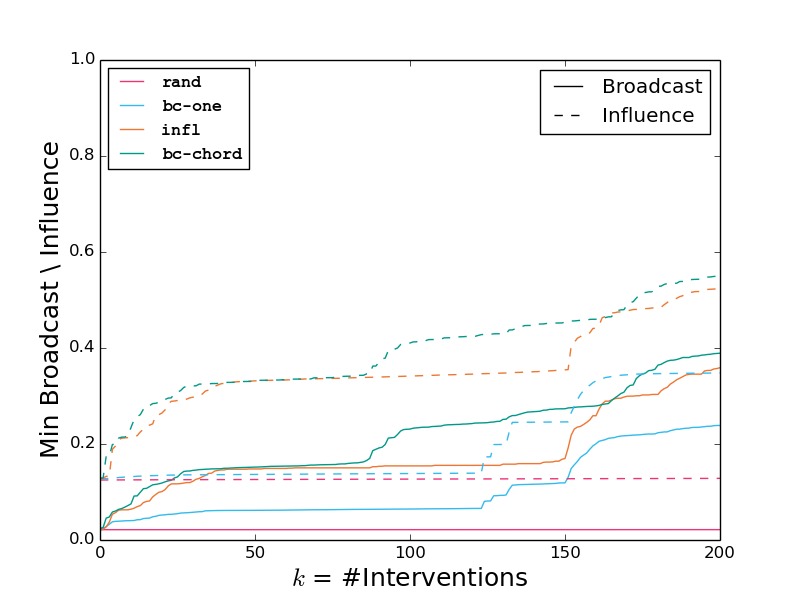}
		\includegraphics[width=.24\columnwidth]{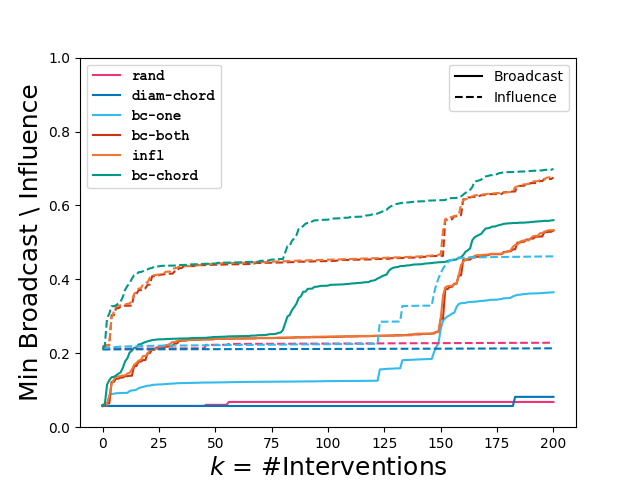}
		 %\caption{Line Plot of Improving Minimum \Minadv{} and \Aveadv{}}
	\end{subfigure}

	\begin{subfigure}[t]{\textwidth}
  		\includegraphics[width=.24\columnwidth]{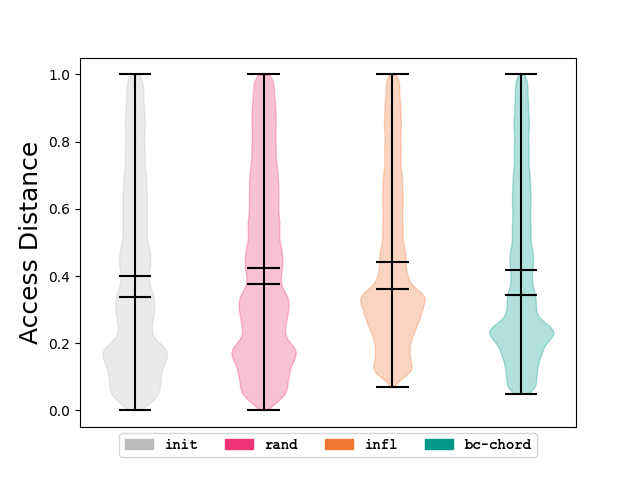}
		\includegraphics[width=.24\columnwidth]{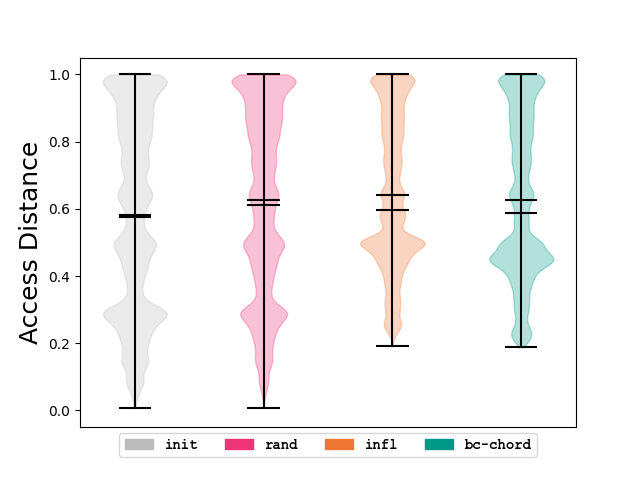}
		\includegraphics[width=.24\columnwidth]{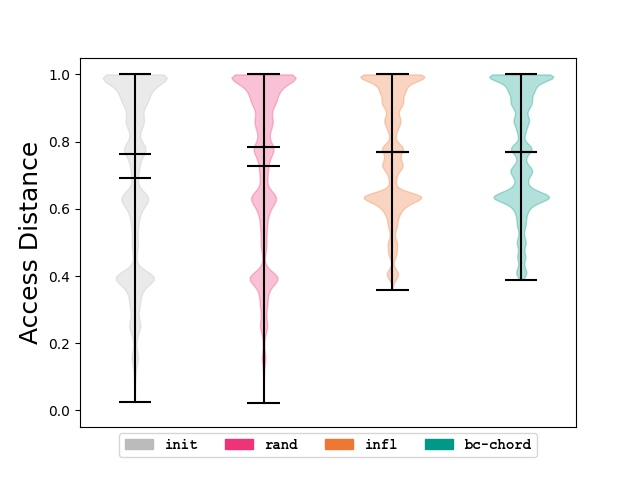}
		\includegraphics[width=.24\columnwidth]{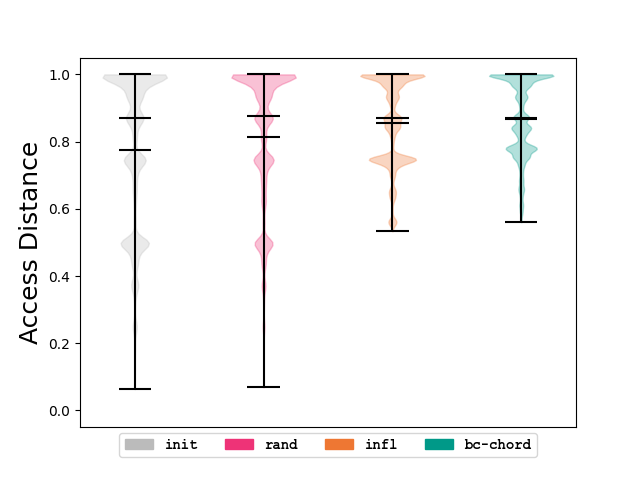}
		 %\caption{Violin Plot of Access Distances}
	\end{subfigure}

	\begin{subfigure}[t]{\textwidth}
  		\includegraphics[width=.24\columnwidth]{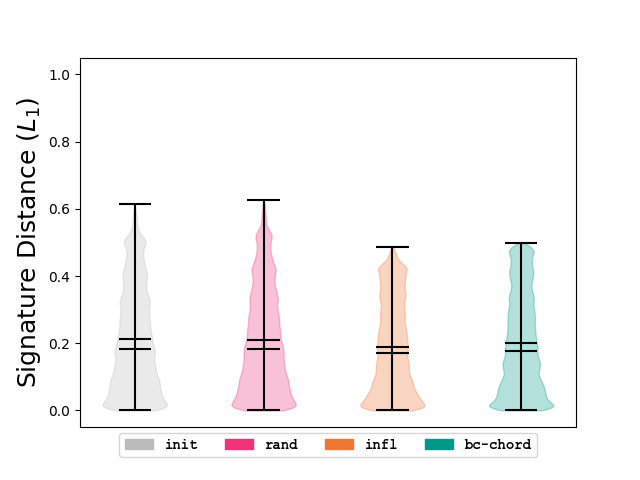}
		\includegraphics[width=.24\columnwidth]{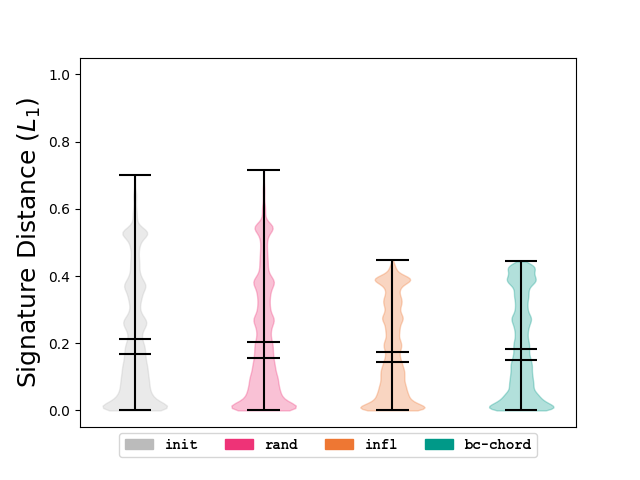}
		\includegraphics[width=.24\columnwidth]{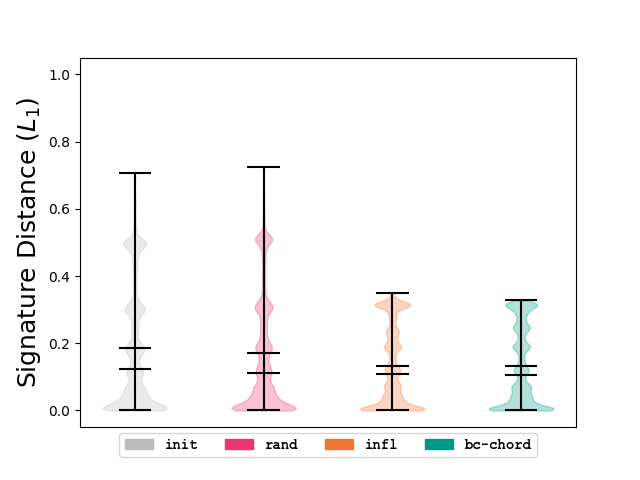}
		\includegraphics[width=.24\columnwidth]{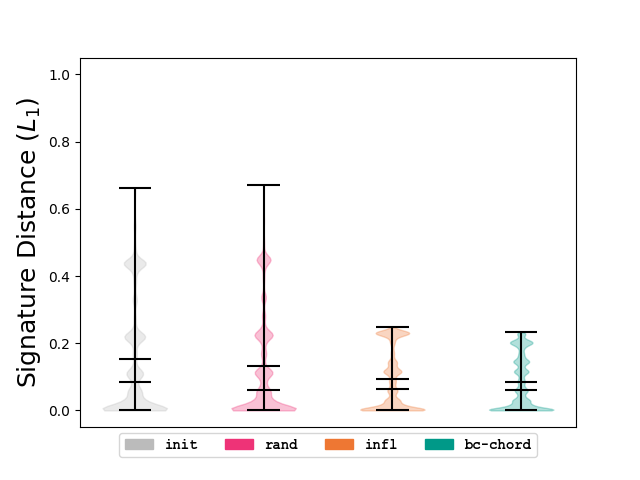}
		 %\caption{Violin plot of Signature Distances.}
	\end{subfigure}

	\caption{Results for Email-Arenas with $\alpha = \{0.2, 0.3, 0.4, 0.5\}$ (L to R).
	At top, we plot improvement in minimum \minadv{} and \aveadv{}; the violin plots show the distribution of pairwise access distances (middle) and $L_1$ signature distances (bottom).}
	\label{fig:complete-one}
\end{figure*}

%BDS Is this where we need to mention unknown complexity?
In this section, we introduce several heuristics for \maxwelfare{} which greedily select
new edges using advantage-based criteria. We employ two basic strategies -- connecting
disadvantaged nodes to a \emph{central} one, and adding a \emph{chord} between two peripheral
nodes. We will compare these with a baseline (\rand) which chooses both ends of each new edge uniformly at random.\looseness-1

We begin by defining the \emph{center} of the network to be the node with maximum \minadv{}.
In our greedy algorithms, we select this node in the un-augmented network and fix it for the duration of the edge selection process. As we iteratively make interventions, it is possible that
a new central node emerges (one with higher broadcast than the selected center). While we could update at
every step, this incurs a high computational cost. In order to evaluate the likelihood and impact of a shifting center, we re-ran the experiments on the three smallest networks and recorded how often the maximum broadcast increased, along with the $L_1$ norm of the access signature difference between initial and new centers. The initial center node remained central more than $99\%$ of the time, and the signature difference was less than $0.01$ in the other $1\%$ of cases. Based on this and the significant computational cost, we choose to fix a center node based on the initial network.

%BDS Removed for space considerations; also heuristics have more intuitive names now!
% \begin{table}
% \vspace*{-0.1in}
% \caption{Overview of Heuristics \label{table:algorithms_set}}
% \vspace{-0.8\baselineskip}
% \begin{tabular}{ p{2cm} p{6cm}}
%  \toprule
%   Name & Endpoints\\
%   \midrule
% {\minpair{}} & $u, v = \argmin_{\substack{v_i, v_j \in V}} \text{\dist{}} (v_i, v_j)$ \\
% {\mincentpair{}} & $u_1 = \argmin_{\substack{v_i \in V}} \text{\minadv{}}(v_i)$, $v = center $\\
% {} & $u_2 = \argmin_{\substack{v_i \in V}} \text{\dist{}}(u_1, v_i)$, $v = center $ \\
% {\mincent{}} & u = $\argmin_{\substack{v_i \in V}} \text{\minadv{}}(v_i)$, $v = center $ \\
% {\minave{}} & u = $\argmin_{\substack{v_i \in V}} \text{\aveadv{}}(v_i)$, $v = center $ \\
% {\diampair{}} & $u, v = \argmax_{\substack{v_i, v_j \in V}} d(v_i, v_j)$ \\
% %BDS: Should we add diamcent here?
%  \bottomrule
% \end{tabular}
% \vspace*{-0.1in}
% \end{table}

Before proceeding to the heuristics, we need two additional observations. First, computing the access distances is known to be $\#P-$hard~\cite{sharpphard10chen}; as such, whenever our strategies use $p_{ij}$,  we rely on simulation to estimate the access distances using Reverse Influence Sampling (RIS)~\cite{Borgs2014MaximizingSI, TangXS14}. Second, greedy heuristics may select a pair of vertices to connect which already have an edge in the graph. When this happens, we select an alternative augmentation in one of two ways: (1) if the heuristic was trying to connect a node $u$ to the \emph{center}, we instead connect $u$ to the node with second-highest \minadv{}, continuing down the \minadv{} order as needed until we find a non-neighbor of $u$; (2) if the heuristic was adding a
\emph{chord} or random edge, we  ``randomly replace an endpoint.'' We can now define our strategies for reducing the access diameter of a network.

\paragraph{\Minadv{}-based Strategies}
To reduce the access diameter of the network we must affect at least one node with minimum \minadv{}.
If $v_i, v_j$ is a pair of nodes so that $p_{ij}$ is minimum, we call them \emph{diameter-defining}.
Our first heuristic \minpair{} finds a diameter-defining pair and adds the edge between them. A natural alternative strategy is to connect one or both of the pair to the \emph{center}; we do this in \mincentpair{} and \mincent{}, respectively. Note that \mincentpair{} adds pairs of edges, and runs for only $\frac{k}{2}$ steps; we constrain $k$ to even values in experiments to ensure fair comparisons.

\paragraph{\Aveadv{}-based Strategies}
Another reasonable approach to improving access in the network is to equalize \aveadv{}. Similar to
\minadv{}, we connect the node with minimum \aveadv{} to the center, and call this heuristic \minave{}.

\paragraph{Diameter-based Strategies}
Finally, we consider a measure that can be computed without simulation, the \emph{diameter} of the underlying network. While the shortest-path distances and access distances may diverge, they are not independent, and creating short paths between nodes will improve their pairwise access. Similar to
\minpair{}, \diampair{} adds an edge between a pair of nodes with maximum $d(u,v)$;
% and \diamcent{} connects them both to a center (adding two edges per round, similar to \mincentpair{}).

\section{Experiments}
\label{sec:expts}

%BDS Figure placement is fragile!
\begin{figure*}
	\centering

	\begin{subfigure}[t]{0.48\textwidth}
		\centering
		\includegraphics[width=.48\columnwidth]{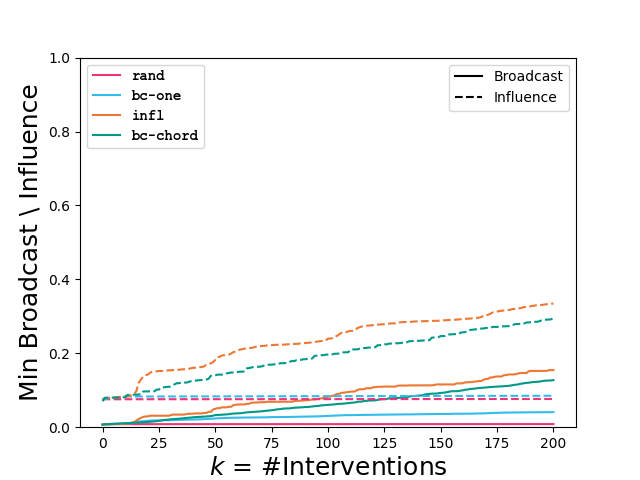}
		\includegraphics[width=.48\columnwidth]{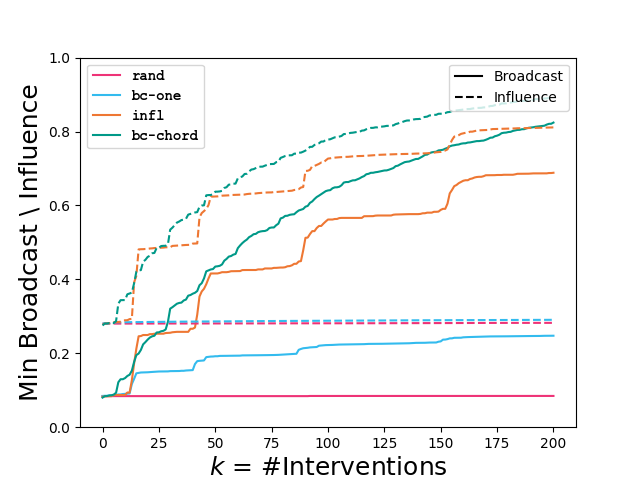}
		\caption{Email-EU (Left: $\alpha=0.1$, Right: $\alpha=0.3$)}
	\end{subfigure}
	~
	\rulesep
	\begin{subfigure}[t]{0.48\textwidth}
		\centering
		\includegraphics[width=.48\columnwidth]{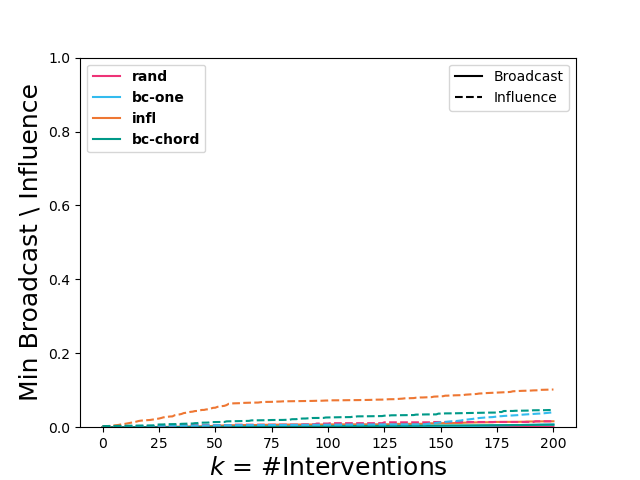}
		\includegraphics[width=.48\columnwidth]{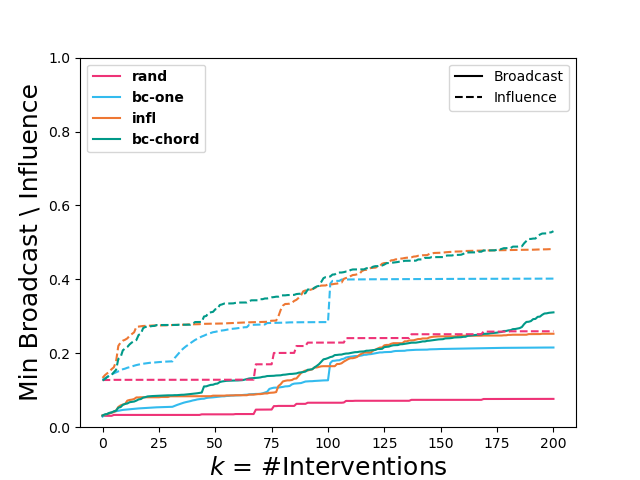}
		\caption{Facebook (Left: $\alpha=0.1$, Right: $\alpha=0.3$)}
	\end{subfigure}

	\begin{subfigure}[t]{0.48\textwidth}
		\centering
		\includegraphics[width=.48\columnwidth]{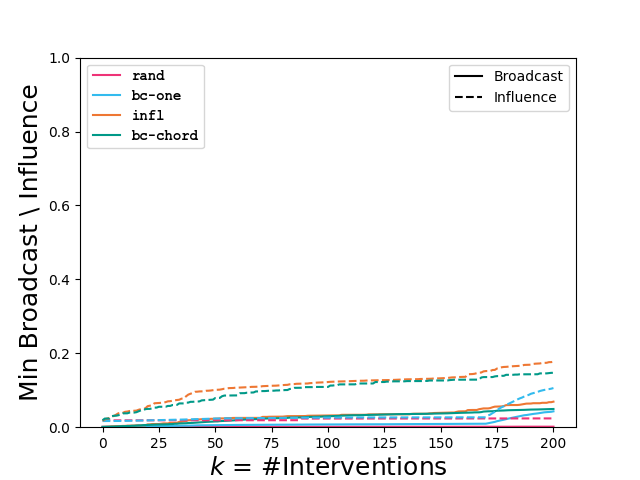}
		\includegraphics[width=.48\columnwidth]{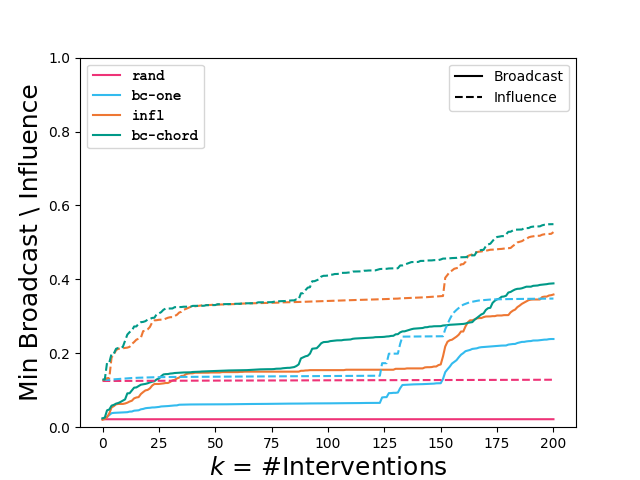}
		\caption{Email-arenas (Left: $\alpha=0.2$, Right: $\alpha=0.4$)}
	\end{subfigure}
	~
	\rulesep
	\begin{subfigure}[t]{0.48\textwidth}
		\centering
		\includegraphics[width=.48\columnwidth]{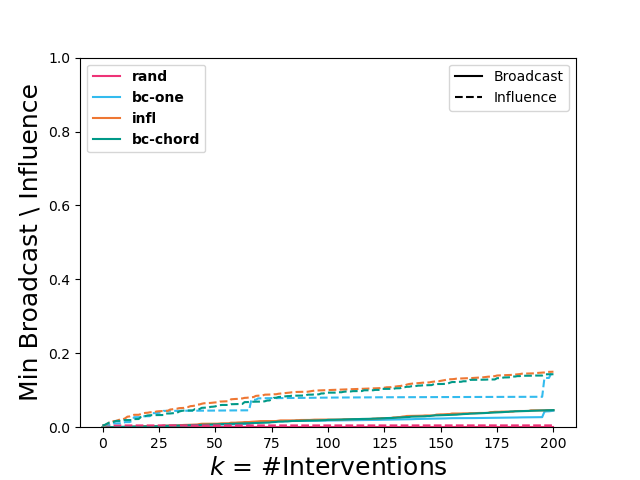}
		\includegraphics[width=.48\columnwidth]{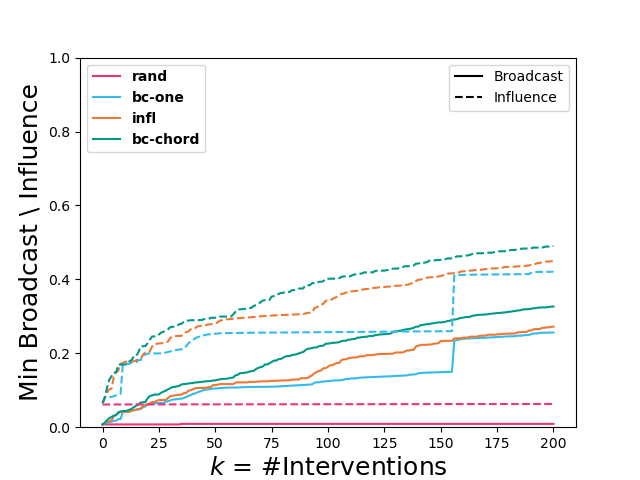}
		\caption{ca-GrQc (Left: $\alpha=0.4$, Right: $\alpha=0.6$)}
	\end{subfigure}

	\begin{subfigure}[t]{0.48\textwidth}
		\centering
		\includegraphics[width=.48\columnwidth]{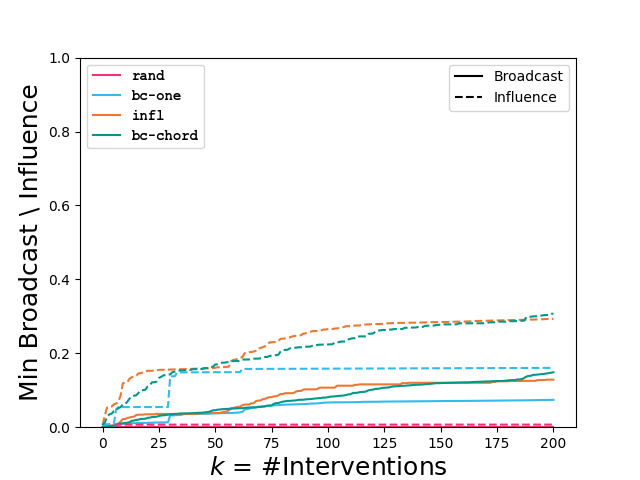}
		\includegraphics[width=.48\columnwidth]{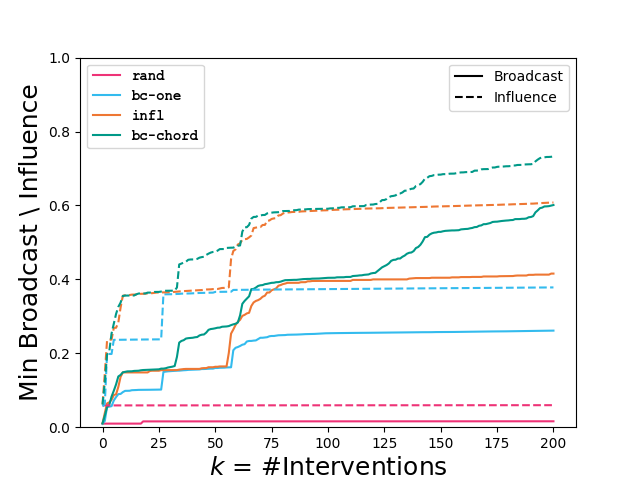}
		\caption{Irvine (Left: $\alpha=0.2$, Right: $\alpha=0.4$)}
	\end{subfigure}
	~
	\rulesep
	\begin{subfigure}[t]{0.48\textwidth}
		\centering
		\includegraphics[width=.48\columnwidth]{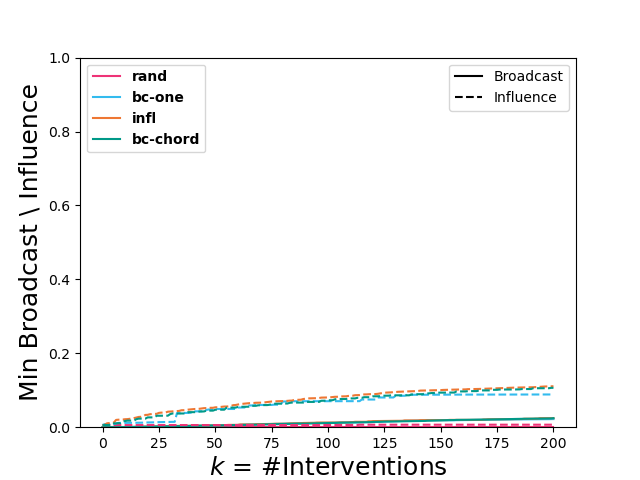}
		\includegraphics[width=.48\columnwidth]{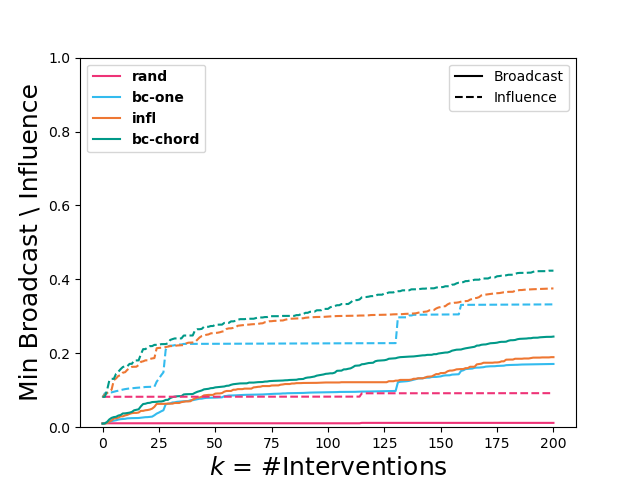}
		\caption{ca-HepTh (Left: $\alpha=0.4$, Right: $\alpha=0.6$)}
	\end{subfigure}

	\caption{For each network, we plot the improvement in min. broadcast and influence for low-moderate- and well-spreading $\alpha$.\looseness-1}
	\label{fig:min-compare}
\end{figure*}

We implemented the heuristics from Section~\ref{sec:heuristics} in C++ and compiled with \texttt{gcc} 8.1.0; all experiments were run on identical hardware equipped with 40 CPUs (Intel Xeon Gold 6230 @ 2.10GHz) and 190 GB of memory, running CentOS Linux release 7.9.2009.
To evaluate the effectiveness of our intervention strategies, we used a corpus of real-world networks sourced from the SNAP~\cite{snapnets} and ICON~\cite{icon} repositories, as described in Table~\ref{table:data_sets}. We treated all data as undirected, and used the largest connected component for each.

As briefly mentioned in Section~\ref{sec:heuristics}, we use Reverse Influence Sampling (RIS)~\cite{Borgs2014MaximizingSI} to estimate access distances; we generate $R = 10,000$ instances per simulation. To evaluate the accuracy, we ran each estimation $10$ times and measured the fluctuations in access distances. In all cases, pairwise accesses varied by less than $0.03$ ($3\%$ of the range), and the average difference was at most $0.004$ ($0.4\%$ of the range). The heuristics \minpair{}, \mincentpair{}, \mincent{}, and \minave{} use RIS, requiring $\O(Rm + Rnk)$ time and $\O(n^2 + Rn)$ space.

In each experiment, we used even values of $k$ from $0$ to $200$, aiming for a practical intervention size relative to the network (less than a tenth of a percent of $|E|$). %(at most $0.06\%$ of $\|E\|$).
In the Independent Cascade model, the spread of information depends on the input parameter $\alpha$ (the probability of transmission along an edge in a time step). For each
network in our corpus, we computed the distribution of access distances for varied $\alpha$ and
selected four (network-specific) values: one each to represent poorly-spreading and well-spreading scenarios, and two in the critical region of moderate spread.

%BDS placement is fragile.
\begin{figure*}
	\centering
	\begin{subfigure}[t]{0.48\textwidth}
		\centering
		\includegraphics[width=.48\columnwidth]{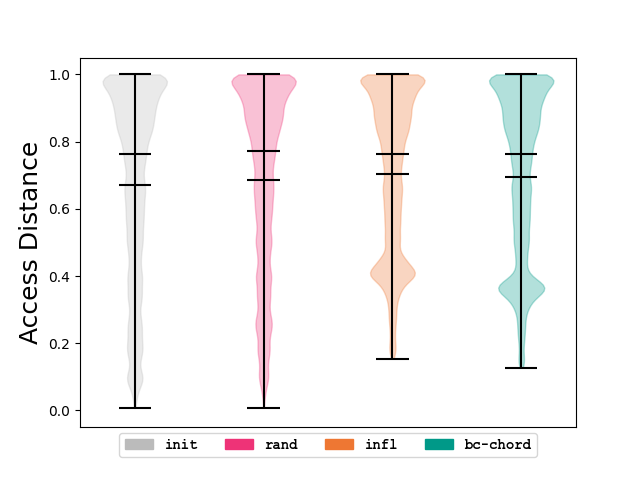}
		\includegraphics[width=.48\columnwidth]{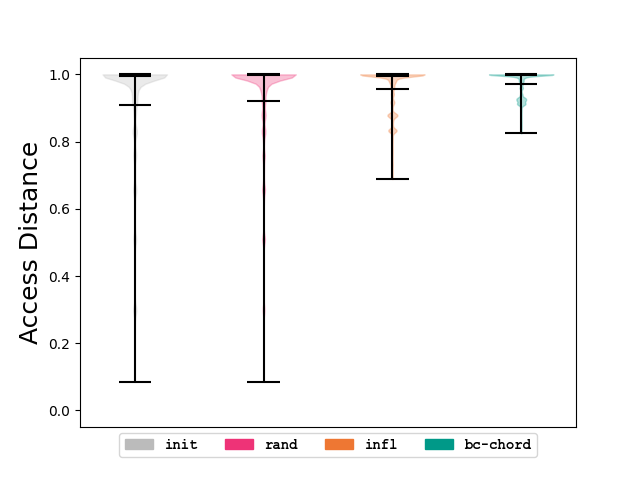}
		\caption{Email-EU (Left: $\alpha=0.1$, Right: $\alpha=0.3$)}
	\end{subfigure}
	~
	\rulesep
	\begin{subfigure}[t]{0.48\textwidth}
		\centering
		\includegraphics[width=.48\columnwidth]{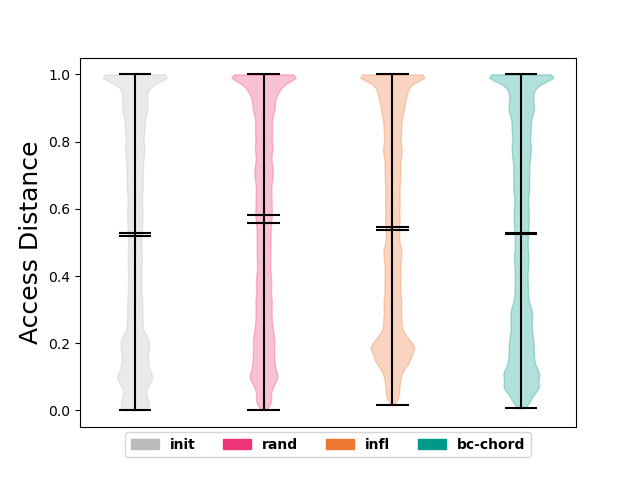}
		\includegraphics[width=.48\columnwidth]{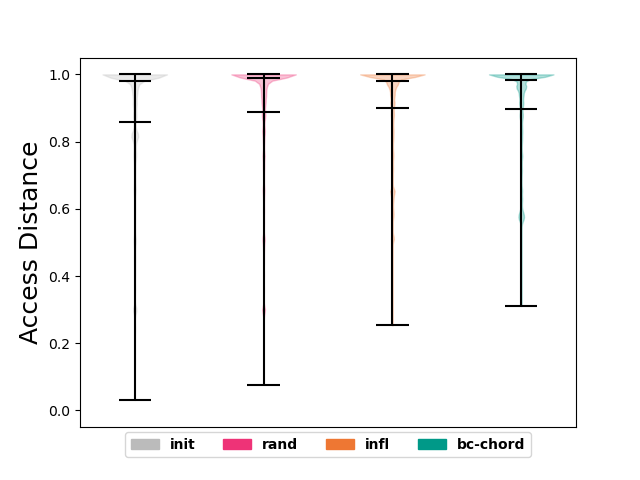}
		\caption{Facebook (Left: $\alpha=0.1$, Right: $\alpha=0.3$)}
	\end{subfigure}

	\begin{subfigure}[t]{0.48\textwidth}
		\centering
		\includegraphics[width=.48\columnwidth]{Edge_MaxMin/Exp/Results/Email-arenas/violin_prob_dist_Email-arenas_20.png}
		\includegraphics[width=.48\columnwidth]{Edge_MaxMin/Exp/Results/Email-arenas/violin_prob_dist_Email-arenas_40.png}
		\caption{Email-arenas (Left: $\alpha=0.2$, Right: $\alpha=0.4$)}
	\end{subfigure}
	~
	\rulesep
	\begin{subfigure}[t]{0.48\textwidth}
		\centering
		\includegraphics[width=.48\columnwidth]{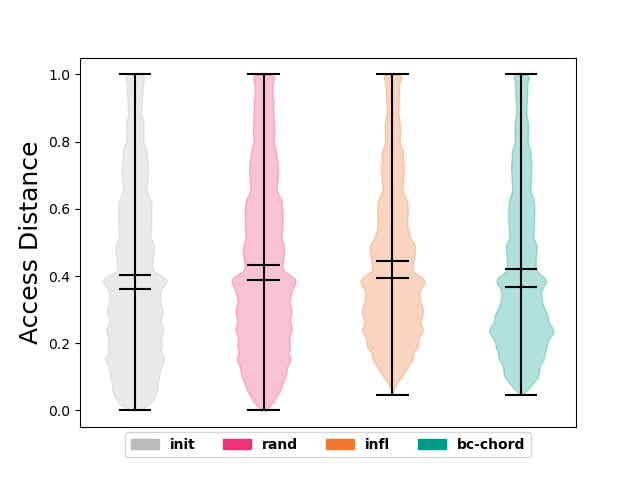}
		\includegraphics[width=.48\columnwidth]{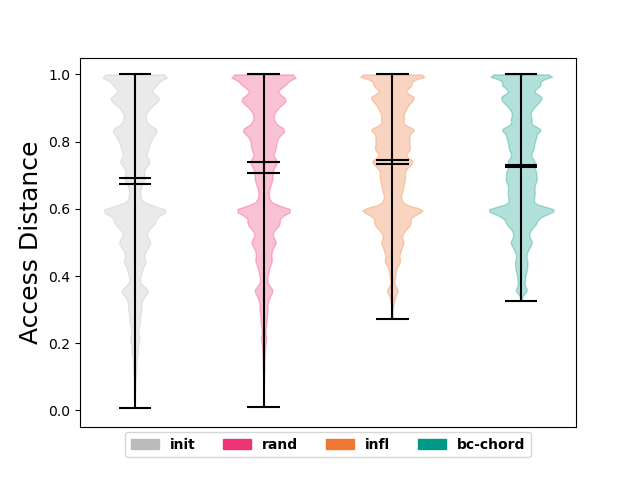}
		\caption{ca-GrQc (Left: $\alpha=0.4$, Right: $\alpha=0.6$)}
	\end{subfigure}

	\begin{subfigure}[t]{0.48\textwidth}
		\centering
		\includegraphics[width=.48\columnwidth]{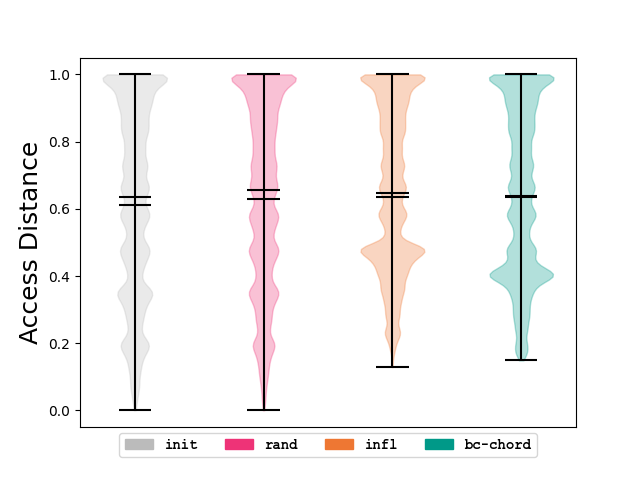}
		\includegraphics[width=.48\columnwidth]{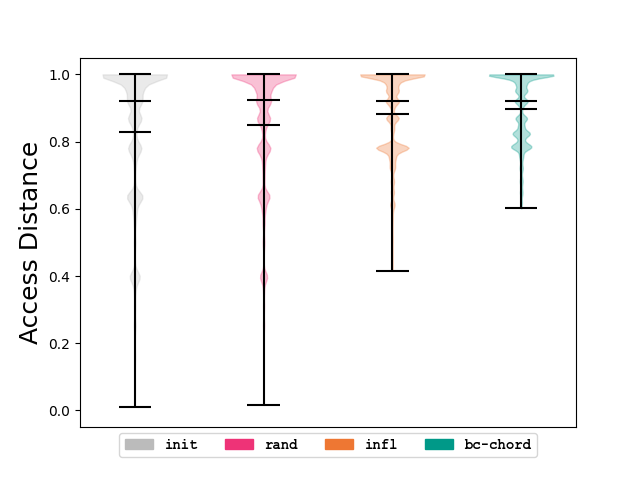}
		\caption{Irvine (Left: $\alpha=0.2$, Right: $\alpha=0.4$)}
	\end{subfigure}
	~
	\rulesep
	\begin{subfigure}[t]{0.48\textwidth}
		\centering
		\includegraphics[width=.48\columnwidth]{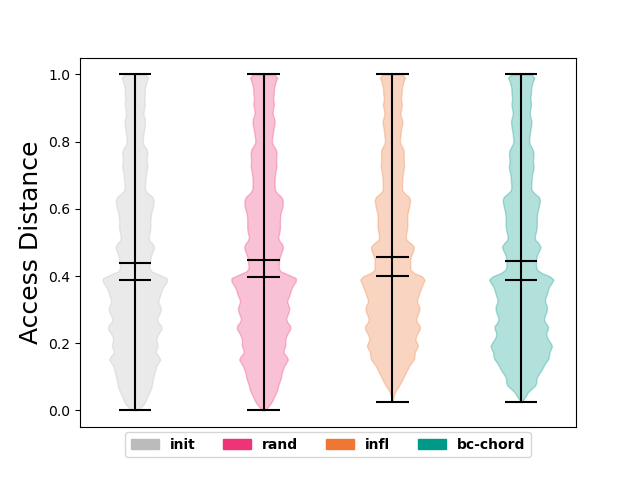}
		\includegraphics[width=.48\columnwidth]{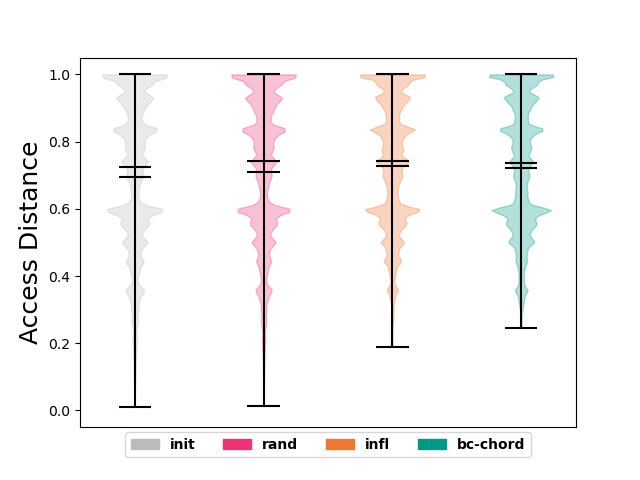}
		\caption{ca-HepTh (Left: $\alpha=0.4$, Right: $\alpha=0.6$)}
	\end{subfigure}

	\caption{For each network, we plot the distribution of pairwise access distances for low-moderate- and well-spreading $\alpha$.\looseness-1}
	\label{fig:shift-mindist}
	\vspace{-0.1em}
\end{figure*}

\subsection{Summary of Experimental Results}
\label{sec:summary}
The primary objective of this work is to intervene in a network to improve access for the most disadvantaged nodes and reduce disparities in advantage by making access signatures more similar.
To assess whether our strategies achieve these goals, we employ several methods for evaluating the outcome of interventions. First, we directly measure the improvement in the minimum values of \minadv{} and \aveadv{} realized in the network. Next, we shift our attention to the access signatures, where we evaluate whether our interventions have increased the similarity among nodes' views of the network using Manhattan distance.  Finally, we consider whether our approaches improve disparity by reducing the advantage gap between the most- and least-privileged nodes.

\begin{table}[b]
\vspace*{-0.1in}
\caption{Summary of Datasets \label{table:data_sets}}
\vspace{-0.8\baselineskip}
\begin{tabular}{ p{2.6cm} p{1cm} p{1cm} p{1.2cm} p{.8cm}}
 \toprule
  Name & Nodes & Edges & Max Deg. & Diam.\\
  \midrule
\href{http://snap.stanford.edu/data/email-EuAll.html}{Email-EU (EU)} & 803 & 24729 & 338 & 5\\
\href{http://konect.cc/networks/arenas-email/}{Email-Arenas (Are)} & 1133 & 5451 & 71 & 8\\
\href{https://toreopsahl.com/datasets/}{Irvine (Irv)} & 1294 & 19026 & 231 & 7\\
\href{https://snap.stanford.edu/data/ego-Facebook.html}{Facebook (Fb)} & 4039 & 88234 & 1045 & 8\\
\href{http://snap.stanford.edu/data/ca-GrQc.html}{ca-GrQc (GrQc)} & 4158 & 13428 & 81 & 17\\
\href{http://snap.stanford.edu/data/ca-HepTh.html}{ca-HepTh (HepTh)} & 8638 & 24827 & 65 & 18\\
 \bottomrule
\end{tabular}
\vspace*{-0.1in}
\end{table}

In Figure~\ref{fig:complete-one}, we present a comprehensive view of all three evaluations for a single network across its four transmission probabilities. From the first row, we observe that the heuristics \minpair{}, \mincentpair{} and \minave{} are most effective at improving \minadv{} and \aveadv{}, with the latter two performing almost identically. Further, \minpair{} surpasses the other approaches as information spreads more easily. These results are qualitatively replicated by the other networks in our corpus (see Section~\ref{sec:maxmin-comparison}). Given this, we restrict our attention to the \minpair{} and \minave{} approaches in subsequent figures, with \minave{} favored over \mincentpair to increase the diversity among our strategies. Further, we note that the behavior with respect to
$\alpha$ remained consistent across all networks, and is well-represented by considering only the low-moderate-spread and well-spreading values of $\alpha$ (2nd and 4th columns). Due to space constraints, plots for the entire corpus (Figures~\ref{fig:min-compare} and ~\ref{fig:shift-mindist}), only show these two transmission probabilities.
In the second row of Figure~\ref{fig:complete-one}, we use violin plots to show the distribution of access distances for all pairs before (\texttt{init}) and after (\rand{}, \minave{}, \minpair{}) intervention. We observe that while randomized augmentation has little effect, both heuristics significantly reduce the maximum pairwise access distance, with \minpair{} again out-performing \minave{} as $\alpha$ increases. While the distributions for other networks vary in initial shape, the pattern of improvement was consistent (see Section~\ref{sec:min-dist}).
Finally, the third row of Figure~\ref{fig:complete-one} illustrates our success in increasing the uniformity among each node's view of the network as measured by reducing the maximum distance between access signatures. Results for other networks are summarized in Section~\ref{sec:min-dist}.

To round out our evaluation, we also computed how our interventions affected the advantage gaps for broadcast, influence, and control, as discussed in Section~\ref{sec:gaps}. For the network featured in Figure~\ref{fig:complete-one}, these results are in the second row of Tables~\ref{table:gaps} and ~\ref{table:controls}. One surprising result was that while the absolute broadcast gap increased, the
relative one decreased. We believe this is caused by interventions increasing access by a larger additive amount for central nodes than peripheral ones. Over the entire corpus, \minpair{} shrank the \minadv{}/\aveadv{} gaps by over $85\%$/$82\%$, respectively.

Overall, we observe that our interventions are most effective when the network is better-connected -- whether because $\alpha$ is higher, or the underlying graph is denser (e.g. in EU and Fb). Additionally, our analysis showed that \mincentpair{} and \minave{} perform almost identically (Figure~\ref{fig:min-compare}), suggesting that the nodes with minimum broadcast and influence may have similar access signatures. To further investigate this phenomenon, we measured the signature difference between the nodes selected by each of these heuristics at each intervention step and found them to be consistently in the bottom $10\%$ of all pairs, with the average falling in the bottom $1\%$. This leads us to hypothesize that the set of least-advantaged nodes with respect to \minadv{} and \aveadv{} are almost identical.

\subsection{Improving Minimum \Minadv{} / \Aveadv{}}
\label{sec:maxmin-comparison}
%BDS this feels repetitive with some of the info in 6.1 now.
The \minadv{} and \aveadv{} measures quantify a node's structural advantage as a function of its signature. Here we evaluate whether edge interventions can improve these measures for the most disadvantaged nodes in the network. Figure~\ref{fig:min-compare} plots the trajectory of the minimum \minadv{} and \aveadv{} as the number of interventions $k$ increases with low-moderate- and well-spreading $\alpha$ for each network in the corpus. We observe that \minave{} and \mincentpair{} consistently show the most improvement for both advantage measures.

%BDS I don't know what this means, but we need a better statistic here.
%The most efficient heuristic increases the advantage of the most disadvantaged node with approximately $\%100$ more effectiveness.
% If bc-one improves by adding X, top heuristics improve by adding 2X.
% if bc-one improves from 0.1 to 0.2 -> top heuristic improves to .3 NOT .4.

%BDS table placement fragile!
\begin{table}
\vspace*{-0.1in}
\caption{Absolute/Relative Advantage Gaps \label{table:gaps}}
\vspace{-0.8\baselineskip}
\centering
\small
\begin{tabular}{ | p{1.5cm} |@{}c@{} | @{}c@{} | }
\hline
(Network, $\alpha$) & Gap & \begin{tabular}{@{}c@{}} Heuristic \\ \hline \begin{tabular}{a | b | d | e} \init & \rand & \minave & \minpair \\\end{tabular} \\ \end{tabular} \\
\hline
(EU, $0.3$) &
    \begin{tabular}{c c} bc \\ \hline infl \end{tabular} &
    \begin{tabular}{a | b | d | e} 0.21/2.49 & 0.21/2.48 & 0.14/0.21 & \textbf{0.08}/\textbf{0.10} \\ \hline 0.67/2.40 & 0.68/2.40 & 0.17/0.20 & \textbf{0.09}/\textbf{0.10} \\ \end{tabular} \\
\hline
(Are, $0.4$) &
    \begin{tabular}{c c} bc \\ \hline infl \end{tabular} &
    \begin{tabular}{a | b | d | e} \textbf{0.13}/5.30 & 0.13/5.95 & 0.24/0.68 & 0.24/\textbf{0.61} \\ \hline 0.71/5.62 & 0.72/5.64 & 0.35/0.66 & \textbf{0.33}/\textbf{0.60} \\ \end{tabular} \\
\hline
(Irv, $0.4$) &
    \begin{tabular}{c c} bc \\ \hline infl \end{tabular} &
    \begin{tabular}{a | b | d | e} \textbf{0.06} /4.83 & 0.08 /5.21 & 0.23 /0.56 & 0.17 /\textbf{0.29} \\ \hline 0.85 /13.8 & 0.86 /14.4 & 0.33 /0.54 & \textbf{0.21} /\textbf{0.29} \\ \end{tabular} \\
\hline
(Fb, $0.3$) &
    \begin{tabular}{c c} bc \\ \hline infl \end{tabular} &
    \begin{tabular}{a | b | d | e} \textbf{0.14} /5.97 & 0.20 /2.63 & 0.26 /0.97 & 0.25 /\textbf{0.79} \\ \hline 0.79 /4.55 & 0.68 /2.61 & 0.47 /1.01 & \textbf{0.42} /\textbf{0.80} \\ \end{tabular} \\
\hline
(GrQc, $0.6$) &
    \begin{tabular}{c c} bc \\ \hline infl \end{tabular} &
    \begin{tabular}{a | b | d | e} \textbf{0.07} /11.6 & 0.07/7.63 & 0.25/0.93 & 0.25/\textbf{0.76} \\ \hline 0.76/12.4 & 0.78/12.4 & 0.41/0.90 & \textbf{0.36}/\textbf{0.74} \\ \end{tabular} \\
\hline
(HepTh, $0.6$) &
    \begin{tabular}{c c} bc \\ \hline infl \end{tabular} &
    \begin{tabular}{a | b | d | e} \textbf{0.09}/9.56 & 0.10/8.33 & 0.25/1.32 & 0.25/\textbf{1.03} \\ \hline 0.75/9.11 & 0.75/8.14 & 0.48/1.27 & \textbf{0.43}/\textbf{1.01} \\ \end{tabular} \\
\hline
\end{tabular}
\vspace*{-0.1in}
\end{table}

\subsection{Making Distances \& Signatures Closer}
\label{sec:min-dist}
One goal of intervention is to increase access for nodes that have the lowest probability of receiving some types of information. In Figure~\ref{fig:shift-mindist}, we plot the distribution of pairwise access distances before and after intervention; we again consider two transmission probabilities (low-moderate-spread and well-spreading) for each of the $6$ networks in the corpus. We observe that while the median value does not move significantly, the lower tail of the distribution gets much shorter and thinner. The amount of improvement increases with $\alpha$, and is more pronounced in the denser networks (EU, Irvine, and Fb).  In some cases, with only $200$ interventions, we are able to increase the minimum pairwise access distance by $0.7$, more doubling the probability of information transmission!
%BDS check that this is accurate.

Another of our objectives is increasing similarity among access signatures so that all nodes have a similar ``view'' of the network. We use the Manhattan distance ($L_1$ metric) to measure the distance between two signatures\footnote{using Euclidean distance ($L_2$) results in similar trends and no qualitative differences}. The third row of Figure~\ref{fig:complete-one} shows violin plots of the distribution of these distances for Email-Arenas; those for other networks are omitted in the interest of space. The maximum signature difference was consistently reduced (at least $43\%$ for well-spreading $\alpha$), and while the median was relatively stable, the tail of the distributions shifted noticeably downward.

\subsection{Measuring the Gap}
\label{sec:gaps}

The final central premise of this work is that improving equity requires reducing access disparities between nodes. To evaluate this, we measure the advantage gap for broadcast and influence, as well as the maximum amount of control achieved in the network (which can be viewed as a gap, since there are always nodes on the periphery with control value essentially zero).

\paragraph{Broadcast/Influence Gaps} We begin by calculating both the absolute ($\max - \min$) and relative ($\frac{\max - \min}{\min}$) advantage gaps for \minadv{} and \aveadv{} on network in the corpus; Table~\ref{table:gaps} shows these when $\alpha$ is well-spreading. As mentioned in Section~\ref{sec:summary}, the absolute broadcast gap often increases with intervention, while the influence gap is typically reduced. However, the relative advantage gap behaves quite differently, consistently decreasing significantly with \minpair{}, yet increases in most cases for \minave{}. This supports our argument that \minave{} may contribute to a rich-get-richer phenomenon by increasing advantage for central nodes, and is an important distinction between two otherwise well-performing heuristics.

\paragraph{Reducing Control} Finally, we consider how our interventions affect \ctradv{}. In Table \ref{table:controls}, we report the maximum values of not only the primary \ctradv{} measure of \cent{} but also the finer-grained pairwise control (\ctradv{}$^{c}_{i}$). Here, we must restrict our analysis to the three smallest networks in our corpus due to the exceptionally high cost of computing control for all nodes (which requires removing each node from the network and re-estimating access distances); we use the same well-spreading $\alpha$ values as in our gap analysis. The results are encouraging, as they show that intervention can increase the independence of nodes in the network when accessing information and prevent better-positioned nodes from having a monopoly over others. It is noteworthy that \minpair{} not only uniformly achieves more than $53\%$ reduction in pairwise control, it never increases the \ctradv{} (whereas \minave{} can cause a $10$-fold jump).

\begin{table}[h]
\vspace*{-0.1in}
\caption{Maximum Control Values \label{table:controls}}
\vspace{-0.8\baselineskip}
\centering
\small
\begin{tabular}{ | p{1.5cm} |@{}c@{} | @{}c@{} | }
\hline
(Network, $\alpha$) & Measure & \begin{tabular}{@{}c@{}} Heuristic \\ \hline \begin{tabular}{ f | g | h | e } \init & \rand & \minave & \minpair \\\end{tabular} \\ \end{tabular} \\
\hline
(EU, $0.3$) &
    \begin{tabular}{c c} \cent{} \\ \hline control$^c_i$ \end{tabular} &
    \begin{tabular}{f | g | h | e} 0.009 & 0.007 & 0.014 & \textbf{0.002} \\ \hline 1.000 & 1.000 & 0.107 & \textbf{0.056} \\ \end{tabular} \\
\hline
(Are, $0.4$) &
    \begin{tabular}{c c} \cent{} \\ \hline control$^c_i$ \end{tabular} &
    \begin{tabular}{f | g | h | e} 0.008 & \textbf{0.006} & 0.112 & 0.008 \\ \hline 1.000 & 1.000 & 0.476 & \textbf{0.464} \\ \end{tabular} \\
\hline
(Irv, $0.4$) &
    \begin{tabular}{c c} \cent{} \\ \hline control$^c_i$  \end{tabular} &
    \begin{tabular}{f | g | h | e} 0.008 & 0.007 & 0.050 & \textbf{0.006} \\ \hline 1.000 & 1.000 & 0.573 & \textbf{0.217} \\ \end{tabular} \\
\hline
\end{tabular}
\vspace*{-0.2in}
\end{table}

\section{Conclusion}
\label{sec:conclusion}

In this work, we propose a novel method for quantifying social capital through the lens of information flow in a network when all nodes have unique, equally-important information to disseminate. We introduce three new measures of structural advantage quantified in terms of network position, argue for intervening through edge augmentation to reduce bias in network structure, and formalize the budgeted intervention problem of \maxwelfare{} for mitigating structural inequity in information access. Finally, we propose heuristic strategies that improve access for the least-advantaged nodes, reduce advantage disparities, and increase the similarity in access signatures.  We perform a case study on a corpus of social networks and demonstrate that our \minpair{} heuristic improves the minimum broadcast \emph{and} influence, dramatically shrink advantage gaps, and reduces variance among access signatures.

Our work is inherently limited by our use of a uniform transmission probability in the Independent Cascade model, and by ignoring the time at which information is received (as we know that early access plays an important role in social capital). Further, the quantification of \ctradv{} is computationally infeasible for large networks, limiting our empirical evaluation.

We leave open many directions for future work, including the adaptation of these ideas to directed networks where access and reach may differ ($p_{ij} \neq p_{ji}$) and optimizing for one may lead to trade-offs for the other. It would also be interesting to adapt this problem to the group fairness setting by defining and optimizing advantage measures on groups. Finally, we note that our measures and strategies can be applied to any probabilistic models of information flow, and may improve many existing diameter-based approaches.

% Another interesting avenue is to apply access-based measure in other settings. In information access setting, these notions better explain the idea of distance rather than the traditionally defined distance defined in graph theory. For example, one application would be applying our problem (minimizing the access diameter of the network) in the settings where the goal is to minimize the diameter instead of the graph to improve communication.

\clearpage

%% The next two lines define the bibliography style to be used, and
%% the bibliography file.
\balance
\bibliography{refs}

%%% -*-BibTeX-*-
%%% Do NOT edit. File created by BibTeX with style
%%% ACM-Reference-Format-Journals [18-Jan-2012].

\begin{thebibliography}{63}

%%% ====================================================================
%%% NOTE TO THE USER: you can override these defaults by providing
%%% customized versions of any of these macros before the \bibliography
%%% command.  Each of them MUST provide its own final punctuation,
%%% except for \shownote{}, \showDOI{}, and \showURL{}.  The latter two
%%% do not use final punctuation, in order to avoid confusing it with
%%% the Web address.
%%%
%%% To suppress output of a particular field, define its macro to expand
%%% to an empty string, or better, \unskip, like this:
%%%
%%% \newcommand{\showDOI}[1]{\unskip}   % LaTeX syntax
%%%
%%% \def \showDOI #1{\unskip}           % plain TeX syntax
%%%
%%% ====================================================================

\ifx \showCODEN    \undefined \def \showCODEN     #1{\unskip}     \fi
\ifx \showDOI      \undefined \def \showDOI       #1{#1}\fi
\ifx \showISBNx    \undefined \def \showISBNx     #1{\unskip}     \fi
\ifx \showISBNxiii \undefined \def \showISBNxiii  #1{\unskip}     \fi
\ifx \showISSN     \undefined \def \showISSN      #1{\unskip}     \fi
\ifx \showLCCN     \undefined \def \showLCCN      #1{\unskip}     \fi
\ifx \shownote     \undefined \def \shownote      #1{#1}          \fi
\ifx \showarticletitle \undefined \def \showarticletitle #1{#1}   \fi
\ifx \showURL      \undefined \def \showURL       {\relax}        \fi
% The following commands are used for tagged output and should be
% invisible to TeX
\providecommand\bibfield[2]{#2}
\providecommand\bibinfo[2]{#2}
\providecommand\natexlab[1]{#1}
\providecommand\showeprint[2][]{arXiv:#2}

\bibitem[Ali et~al\mbox{.}(2019)]%
        {ali2019fairness}
\bibfield{author}{\bibinfo{person}{Junaid Ali}, \bibinfo{person}{Mahmoudreza
  Babaei}, \bibinfo{person}{Abhijnan Chakraborty}, \bibinfo{person}{Baharan
  Mirzasoleiman}, \bibinfo{person}{Krishna~P. Gummadi}, {and}
  \bibinfo{person}{Adish Singla}.} \bibinfo{year}{2019}\natexlab{}.
\newblock \bibinfo{title}{On the Fairness of Time-Critical Influence
  Maximization in Social Networks}.
\newblock
\newblock


\bibitem[Ball and Newman(2013)]%
        {ball2012}
\bibfield{author}{\bibinfo{person}{Brian Ball} {and} \bibinfo{person}{M.E.J.
  Newman}.} \bibinfo{year}{2013}\natexlab{}.
\newblock \showarticletitle{Friendship networks and social status}.
\newblock \bibinfo{journal}{\emph{Network Science}} \bibinfo{volume}{1},
  \bibinfo{number}{1} (\bibinfo{year}{2013}).
\newblock


\bibitem[Becker et~al\mbox{.}(2020)]%
        {Becker2020}
\bibfield{author}{\bibinfo{person}{Ruben Becker}, \bibinfo{person}{Federico
  Corò}, \bibinfo{person}{Gianlorenzo D’Angelo}, {and} \bibinfo{person}{Hugo
  Gilbert}.} \bibinfo{year}{2020}\natexlab{}.
\newblock \showarticletitle{Balancing Spreads of Influence in a Social
  Network}.
\newblock \bibinfo{journal}{\emph{Proceedings of the AAAI Conference on
  Artificial Intelligence}} \bibinfo{volume}{34}, \bibinfo{number}{01}
  (\bibinfo{date}{Apr.} \bibinfo{year}{2020}), \bibinfo{pages}{3--10}.
\newblock


\bibitem[Becker et~al\mbox{.}(2021)]%
        {becker2021}
\bibfield{author}{\bibinfo{person}{Ruben Becker}, \bibinfo{person}{Gianlorenzo
  D’Angelo}, \bibinfo{person}{Sajjad Ghobadi}, {and} \bibinfo{person}{Hugo
  Gilbert}.} \bibinfo{year}{2021}\natexlab{}.
\newblock \showarticletitle{Fairness in Influence Maximization through
  Randomization}.
\newblock \bibinfo{journal}{\emph{Proceedings of the AAAI Conference on
  Artificial Intelligence}} \bibinfo{number}{17} (\bibinfo{date}{May}
  \bibinfo{year}{2021}), \bibinfo{pages}{14684--14692}.
\newblock


\bibitem[Beilinson et~al\mbox{.}(2020)]%
        {beilinson2020clustering}
\bibfield{author}{\bibinfo{person}{Hannah~C. Beilinson},
  \bibinfo{person}{Nasanbayar Ulzii-Orshikh}, \bibinfo{person}{Ashkan
  Bashardoust}, \bibinfo{person}{Sorelle~A. Friedler},
  \bibinfo{person}{Carlos~E. Scheidegger}, {and} \bibinfo{person}{Suresh
  Venkatasubramanian}.} \bibinfo{year}{2020}\natexlab{}.
\newblock \showarticletitle{Clustering via Information Access in a Network}.
\newblock \bibinfo{journal}{\emph{arXiv}}  \bibinfo{volume}{abs/2010.12611}
  (\bibinfo{year}{2020}).
\newblock


\bibitem[Borgs et~al\mbox{.}(2014)]%
        {Borgs2014MaximizingSI}
\bibfield{author}{\bibinfo{person}{Christian Borgs}, \bibinfo{person}{Michael
  Brautbar}, \bibinfo{person}{Jennifer Chayes}, {and} \bibinfo{person}{Brendan
  Lucier}.} \bibinfo{year}{2014}\natexlab{}.
\newblock \showarticletitle{Maximizing Social Influence in Nearly Optimal
  Time}. In \bibinfo{booktitle}{\emph{Proceedings of the 2014 Annual ACM-SIAM
  Symposium on Discrete Algorithms (SODA)}}. \bibinfo{pages}{946--957}.
\newblock


\bibitem[Boyd et~al\mbox{.}(2014)]%
        {boyd2014networked}
\bibfield{author}{\bibinfo{person}{Danah Boyd}, \bibinfo{person}{Karen Levy},
  {and} \bibinfo{person}{Alice Marwick}.} \bibinfo{year}{2014}\natexlab{}.
\newblock \showarticletitle{The networked nature of algorithmic
  discrimination}.
\newblock \bibinfo{journal}{\emph{Data and Discrimination: Collected Essays.
  Open Technology Institute}} (\bibinfo{year}{2014}).
\newblock


\bibitem[Brass(1992)]%
        {brass1992power}
\bibfield{author}{\bibinfo{person}{Daniel~J. Brass}.}
  \bibinfo{year}{1992}\natexlab{}.
\newblock \showarticletitle{Power in Organizations: A Social Network
  Perspective}.
\newblock \bibinfo{journal}{\emph{Research in Politics and Society}}
  \bibinfo{volume}{4} (\bibinfo{year}{1992}), \bibinfo{pages}{295--323}.
\newblock
Issue 1.


\bibitem[Burt(2004)]%
        {burt2004holes}
\bibfield{author}{\bibinfo{person}{Ronald S. Burt}.}
  \bibinfo{year}{2004}\natexlab{}.
\newblock \showarticletitle{Structural Holes and Good Ideas}.
\newblock \bibinfo{journal}{\emph{Amer. J. Sociology}} \bibinfo{volume}{110},
  \bibinfo{number}{2} (\bibinfo{year}{2004}), \bibinfo{pages}{349--399}.
\newblock


\bibitem[Burt(1987)]%
        {burt1987cohesion}
\bibfield{author}{\bibinfo{person}{Ronald~S. Burt}.}
  \bibinfo{year}{1987}\natexlab{}.
\newblock \showarticletitle{Social Contagion and Innovation: Cohesion versus
  Structural Equivalence}.
\newblock \bibinfo{journal}{\emph{Amer. J. Sociology}} \bibinfo{volume}{92},
  \bibinfo{number}{6} (\bibinfo{year}{1987}), \bibinfo{pages}{1287--1335}.
\newblock


\bibitem[Burt(1999)]%
        {burt1999opinion}
\bibfield{author}{\bibinfo{person}{Ronald~S. Burt}.}
  \bibinfo{year}{1999}\natexlab{}.
\newblock \showarticletitle{The Social Capital of Opinion Leaders}.
\newblock \bibinfo{journal}{\emph{The Annals of the American Academy of
  Political and Social Science}}  \bibinfo{volume}{566} (\bibinfo{year}{1999}),
  \bibinfo{pages}{37--54}.
\newblock


\bibitem[Burt(2000)]%
        {burt2000socialcapital}
\bibfield{author}{\bibinfo{person}{Ronald~S. Burt}.}
  \bibinfo{year}{2000}\natexlab{}.
\newblock \showarticletitle{The Network Structure Of Social Capital}.
\newblock \bibinfo{journal}{\emph{Research in Organizational Behavior}}
  \bibinfo{volume}{22} (\bibinfo{year}{2000}), \bibinfo{pages}{345--423}.
\newblock
\showISSN{0191-3085}


\bibitem[Burt et~al\mbox{.}(2000)]%
        {Burt2000BridgeCount}
\bibfield{author}{\bibinfo{person}{Ronald~S. Burt}, \bibinfo{person}{Robin~M.
  Hogarth}, {and} \bibinfo{person}{Claude Michaud}.}
  \bibinfo{year}{2000}\natexlab{}.
\newblock \showarticletitle{The Social Capital of French and American
  Managers}.
\newblock \bibinfo{journal}{\emph{Organization Science}}  \bibinfo{volume}{11}
  (\bibinfo{year}{2000}), \bibinfo{pages}{123--147}.
\newblock


\bibitem[Burt et~al\mbox{.}(2013)]%
        {burt2013advantage}
\bibfield{author}{\bibinfo{person}{Ronald~S. Burt}, \bibinfo{person}{Martin
  Kilduff}, {and} \bibinfo{person}{Stefano Tasselli}.}
  \bibinfo{year}{2013}\natexlab{}.
\newblock \showarticletitle{Social Network Analysis: Foundations and Frontiers
  on Advantage}.
\newblock \bibinfo{journal}{\emph{Annual Review of Psychology}}
  \bibinfo{volume}{64}, \bibinfo{number}{1} (\bibinfo{year}{2013}),
  \bibinfo{pages}{527--547}.
\newblock


\bibitem[Campbell et~al\mbox{.}(1986)]%
        {campbell1986}
\bibfield{author}{\bibinfo{person}{Karen~E. Campbell},
  \bibinfo{person}{Peter~V. Marsden}, {and} \bibinfo{person}{Jeanne~S.
  Hurlbert}.} \bibinfo{year}{1986}\natexlab{}.
\newblock \showarticletitle{Social Resources and Socioeconomic Status}.
\newblock \bibinfo{journal}{\emph{Social Networks}} \bibinfo{volume}{8},
  \bibinfo{number}{1} (\bibinfo{year}{1986}), \bibinfo{pages}{97--117}.
\newblock


\bibitem[Chen et~al\mbox{.}(2010)]%
        {sharpphard10chen}
\bibfield{author}{\bibinfo{person}{Wei Chen}, \bibinfo{person}{Chi Wang}, {and}
  \bibinfo{person}{Yajun Wang}.} \bibinfo{year}{2010}\natexlab{}.
\newblock \showarticletitle{Scalable Influence Maximization for Prevalent Viral
  Marketing in Large-Scale Social Networks}.
\newblock \bibinfo{journal}{\emph{Proceedings of the ACM SIGKDD International
  Conference on Knowledge Discovery and Data Mining}},
  \bibinfo{pages}{1029--1038}.
\newblock


\bibitem[Clauset et~al\mbox{.}(2015)]%
        {clauset2015}
\bibfield{author}{\bibinfo{person}{Aaron Clauset}, \bibinfo{person}{Samuel
  Arbesman}, {and} \bibinfo{person}{Daniel~B. Larremore}.}
  \bibinfo{year}{2015}\natexlab{}.
\newblock \showarticletitle{Systematic inequality and hierarchy in faculty
  hiring networks}.
\newblock \bibinfo{journal}{\emph{Science Advances}} \bibinfo{volume}{1},
  \bibinfo{number}{1} (\bibinfo{year}{2015}), \bibinfo{pages}{e1400005}.
\newblock


\bibitem[Clauset et~al\mbox{.}(2016)]%
        {icon}
\bibfield{author}{\bibinfo{person}{Aaron Clauset}, \bibinfo{person}{Ellen
  Tucker}, {and} \bibinfo{person}{Matthias Sainz}.}
  \bibinfo{year}{2016}\natexlab{}.
\newblock \bibinfo{title}{The Colorado Index of Complex Networks.}
\newblock
\newblock
\urldef\tempurl%
\url{https://icon.colorado.edu/}
\showURL{%
\tempurl}


\bibitem[Coleman(1988)]%
        {coleman1988social}
\bibfield{author}{\bibinfo{person}{James~S. Coleman}.}
  \bibinfo{year}{1988}\natexlab{}.
\newblock \showarticletitle{Social Capital in the Creation of Human Capital}.
\newblock \bibinfo{journal}{\emph{Amer. J. Sociology}}  \bibinfo{volume}{94}
  (\bibinfo{year}{1988}), \bibinfo{pages}{S95--S120}.
\newblock


\bibitem[Coleman et~al\mbox{.}(1966)]%
        {colemanmedical}
\bibfield{author}{\bibinfo{person}{James~S. Coleman}, \bibinfo{person}{Elihu
  Katz}, {and} \bibinfo{person}{Herbert Menzel}.}
  \bibinfo{year}{1966}\natexlab{}.
\newblock \bibinfo{booktitle}{\emph{Medical Innovation: A diffusion study}}.
\newblock \bibinfo{publisher}{Bobbs-Merril}, \bibinfo{address}{New York}.
\newblock


\bibitem[danah boyd(2021)]%
        {boyd2021knit}
\bibfield{author}{\bibinfo{person}{danah boyd}.}
  \bibinfo{year}{2021}\natexlab{}.
\newblock \bibinfo{title}{Knitting a Healthy Social Fabric.}
\newblock
\newblock
\urldef\tempurl%
\url{https://zephoria.medium.com/knitting-a-healthy-social-fabric-86105cb92c1c}
\showURL{%
\tempurl}


\bibitem[D'Angelo et~al\mbox{.}(2019)]%
        {dangelo2019recom}
\bibfield{author}{\bibinfo{person}{Gianlorenzo D'Angelo},
  \bibinfo{person}{Lorenzo Severini}, {and} \bibinfo{person}{Yllka Velaj}.}
  \bibinfo{year}{2019}\natexlab{}.
\newblock \showarticletitle{Recommending links through influence maximization}.
\newblock \bibinfo{journal}{\emph{Theoretical Computer Science}}
  \bibinfo{volume}{764} (\bibinfo{year}{2019}), \bibinfo{pages}{30--41}.
\newblock


\bibitem[Demaine and Zadimoghaddam(2010)]%
        {demaine2010diam}
\bibfield{author}{\bibinfo{person}{Erik Demaine} {and} \bibinfo{person}{Morteza
  Zadimoghaddam}.} \bibinfo{year}{2010}\natexlab{}.
\newblock \showarticletitle{Minimizing the Diameter of a Network Using Shortcut
  Edges}. In \bibinfo{booktitle}{\emph{Algorithm Theory - SWAT 2010}}.
  \bibinfo{publisher}{Springer Berlin Heidelberg}, \bibinfo{pages}{420--431}.
\newblock


\bibitem[Domingos and Richardson(2001)]%
        {domingos01mining}
\bibfield{author}{\bibinfo{person}{Pedro Domingos} {and} \bibinfo{person}{Matt
  Richardson}.} \bibinfo{year}{2001}\natexlab{}.
\newblock \showarticletitle{Mining the Network Value of Customers}.
\newblock \bibinfo{journal}{\emph{Proceedings of the Seventh International
  Conference on Knowledge Discovery and Data Mining}} (\bibinfo{year}{2001}),
  \bibinfo{pages}{57–66}.
\newblock


\bibitem[Fish et~al\mbox{.}(2019)]%
        {Fish19Gaps}
\bibfield{author}{\bibinfo{person}{Benjamin Fish}, \bibinfo{person}{Ashkan
  Bashardoust}, \bibinfo{person}{danah boyd}, \bibinfo{person}{Sorelle
  Friedler}, \bibinfo{person}{Carlos Scheidegger}, {and}
  \bibinfo{person}{Suresh Venkatasubramanian}.}
  \bibinfo{year}{2019}\natexlab{}.
\newblock \showarticletitle{Gaps in Information Access in Social Networks?}. In
  \bibinfo{booktitle}{\emph{The World Wide Web Conference}}.
  \bibinfo{publisher}{Association for Computing Machinery},
  \bibinfo{address}{New York, NY, USA}, \bibinfo{pages}{480–490}.
\newblock


\bibitem[Freeman(1977)]%
        {FreemanBetween}
\bibfield{author}{\bibinfo{person}{Linton~C. Freeman}.}
  \bibinfo{year}{1977}\natexlab{}.
\newblock \showarticletitle{A Set of Measures of Centrality Based on
  Betweenness}.
\newblock \bibinfo{journal}{\emph{Sociometry}} \bibinfo{volume}{40},
  \bibinfo{number}{1} (\bibinfo{year}{1977}), \bibinfo{pages}{35--41}.
\newblock


\bibitem[Fung et~al\mbox{.}(2015)]%
        {socialmediapublichealth}
\bibfield{author}{\bibinfo{person}{I.~C. Fung}, \bibinfo{person}{Z.~T. Tse},
  {and} \bibinfo{person}{K.~W. Fu}.} \bibinfo{year}{2015}\natexlab{}.
\newblock \showarticletitle{The use of social media in public health
  surveillance}.
\newblock \bibinfo{journal}{\emph{Western Pac Surveill Response Journal}}
  (\bibinfo{date}{6} \bibinfo{year}{2015}).
\newblock


\bibitem[Geroski and Mazzucato(2002)]%
        {geroski2002learning}
\bibfield{author}{\bibinfo{person}{Paul~A. Geroski} {and}
  \bibinfo{person}{Mariana Mazzucato}.} \bibinfo{year}{2002}\natexlab{}.
\newblock \showarticletitle{Learning and the sources of corporate growth}.
\newblock \bibinfo{journal}{\emph{Industrial and Corporate Change}}
  \bibinfo{volume}{11}, \bibinfo{number}{4} (\bibinfo{year}{2002}),
  \bibinfo{pages}{623--644}.
\newblock


\bibitem[Granovetter(1973)]%
        {granovetter77strength}
\bibfield{author}{\bibinfo{person}{Mark~S Granovetter}.}
  \bibinfo{year}{1973}\natexlab{}.
\newblock \showarticletitle{The strength of weak ties}.
\newblock \bibinfo{journal}{\emph{The American Journal of Sociology}}
  \bibinfo{volume}{78}, \bibinfo{number}{6} (\bibinfo{year}{1973}),
  \bibinfo{pages}{1360--1380}.
\newblock


\bibitem[Jackson(2019)]%
        {jackson19human}
\bibfield{author}{\bibinfo{person}{Matthew Jackson}.}
  \bibinfo{year}{2019}\natexlab{}.
\newblock \bibinfo{booktitle}{\emph{The Human Network: How Your Social Position
  Determines Your Power, Beliefs, and Behaviors}}.
\newblock \bibinfo{publisher}{Knopf Doubleday Publishing Group}.
\newblock
\showISBNx{1101871431}


\bibitem[Jalali et~al\mbox{.}(2020)]%
        {jalali2020unfairness}
\bibfield{author}{\bibinfo{person}{Zeinab~S. Jalali}, \bibinfo{person}{Weixiang
  Wang}, \bibinfo{person}{Myunghwan Kim}, \bibinfo{person}{Hema Raghavan},
  {and} \bibinfo{person}{Sucheta Soundarajan}.}
  \bibinfo{year}{2020}\natexlab{}.
\newblock \showarticletitle{On the information unfairness of social networks}.
  In \bibinfo{booktitle}{\emph{Proceedings of the 2020 SIAM International
  Conference on Data Mining (SDM)}}. \bibinfo{pages}{613--521}.
\newblock


\bibitem[Karimi et~al\mbox{.}(2018)]%
        {karimi2018}
\bibfield{author}{\bibinfo{person}{Fariba Karimi}, \bibinfo{person}{Mathieu
  Génois}, \bibinfo{person}{Claudia Wagner}, \bibinfo{person}{Philipp Singer},
  {and} \bibinfo{person}{Markus Strohmaier}.} \bibinfo{year}{2018}\natexlab{}.
\newblock \showarticletitle{Homophily influences ranking of minorities in
  social networks}.
\newblock \bibinfo{journal}{\emph{Scientific Reports}}  \bibinfo{volume}{8}
  (\bibinfo{date}{07} \bibinfo{year}{2018}).
\newblock


\bibitem[Katz and Lazarsfeld(1966)]%
        {katz1966personal}
\bibfield{author}{\bibinfo{person}{E. Katz} {and} \bibinfo{person}{P.F.
  Lazarsfeld}.} \bibinfo{year}{1966}\natexlab{}.
\newblock \bibinfo{booktitle}{\emph{Personal Influence: the Part Played by
  People in the Flow of Mass Communications}}.
\newblock \bibinfo{publisher}{Free Press}.
\newblock
\showISBNx{9781412830706}


\bibitem[Kempe et~al\mbox{.}(2003)]%
        {kempe03maximizing}
\bibfield{author}{\bibinfo{person}{David Kempe}, \bibinfo{person}{Jon
  Kleinberg}, {and} \bibinfo{person}{{\'{E}}va Tardos}.}
  \bibinfo{year}{2003}\natexlab{}.
\newblock \showarticletitle{Maximizing the Spread of Influence through a Social
  Network}.
\newblock \bibinfo{journal}{\emph{Proceedings of the ACM SIGKDD International
  Conference on Knowledge Discovery and Data Mining}} (\bibinfo{year}{2003}),
  \bibinfo{pages}{137--146}.
\newblock


\bibitem[Knoke and Burt(1983)]%
        {prominence}
\bibfield{author}{\bibinfo{person}{David Knoke} {and} \bibinfo{person}{Ronald~S
  Burt}.} \bibinfo{year}{1983}\natexlab{}.
\newblock \showarticletitle{Prominence}.
\newblock \bibinfo{journal}{\emph{Applied Network Analysis}}
  (\bibinfo{year}{1983}), \bibinfo{pages}{195--222}.
\newblock


\bibitem[Kogan et~al\mbox{.}(2015)]%
        {Kogan2015Think}
\bibfield{author}{\bibinfo{person}{Marina Kogan}, \bibinfo{person}{Leysia
  Palen}, {and} \bibinfo{person}{Kenneth~Mark Anderson}.}
  \bibinfo{year}{2015}\natexlab{}.
\newblock \showarticletitle{Think Local, Retweet Global: Retweeting by the
  Geographically-Vulnerable during Hurricane Sandy}.
\newblock \bibinfo{journal}{\emph{Proceedings of the 18th ACM Conference on
  Computer Supported Cooperative Work \& Social Computing}}
  (\bibinfo{year}{2015}).
\newblock


\bibitem[Leskovec et~al\mbox{.}(2008)]%
        {leskovec2008}
\bibfield{author}{\bibinfo{person}{Jure Leskovec}, \bibinfo{person}{Lars
  Backstrom}, \bibinfo{person}{Ravi Kumar}, {and} \bibinfo{person}{Andrew
  Tomkins}.} \bibinfo{year}{2008}\natexlab{}.
\newblock \showarticletitle{Microscopic evolution of social networks}.
\newblock \bibinfo{journal}{\emph{Proceedings of the ACM SIGKDD International
  Conference on Knowledge Discovery and Data Mining}},
  \bibinfo{pages}{462--470}.
\newblock


\bibitem[Leskovec et~al\mbox{.}(2010)]%
        {leskovec2010}
\bibfield{author}{\bibinfo{person}{Jure Leskovec}, \bibinfo{person}{Deepayan
  Chakrabarti}, \bibinfo{person}{Jon Kleinberg}, \bibinfo{person}{Christos
  Faloutsos}, {and} \bibinfo{person}{Zoubin Ghahramani}.}
  \bibinfo{year}{2010}\natexlab{}.
\newblock \showarticletitle{Kronecker Graphs: An Approach to Modeling
  Networks}.
\newblock \bibinfo{journal}{\emph{J. Mach. Learn. Res.}}  \bibinfo{volume}{11}
  (\bibinfo{year}{2010}), \bibinfo{pages}{985–1042}.
\newblock


\bibitem[Leskovec and Krevl(2014)]%
        {snapnets}
\bibfield{author}{\bibinfo{person}{Jure Leskovec} {and} \bibinfo{person}{Andrej
  Krevl}.} \bibinfo{year}{2014}\natexlab{}.
\newblock \bibinfo{title}{{SNAP Datasets}: {Stanford} Large Network Dataset
  Collection}.
\newblock \bibinfo{howpublished}{\url{http://snap.stanford.edu/data}}.
\newblock


\bibitem[Li et~al\mbox{.}(2018)]%
        {li2018influence}
\bibfield{author}{\bibinfo{person}{Yuchen Li}, \bibinfo{person}{Ju Fan},
  \bibinfo{person}{Yanhao Wang}, {and} \bibinfo{person}{Kian-Lee Tan}.}
  \bibinfo{year}{2018}\natexlab{}.
\newblock \showarticletitle{Influence maximization on social graphs: A survey}.
\newblock \bibinfo{journal}{\emph{IEEE Transactions on Knowledge and Data
  Engineering}} \bibinfo{volume}{30}, \bibinfo{number}{10}
  (\bibinfo{year}{2018}), \bibinfo{pages}{1852--1872}.
\newblock


\bibitem[Lin(1999)]%
        {lin1999}
\bibfield{author}{\bibinfo{person}{Nan Lin}.} \bibinfo{year}{1999}\natexlab{}.
\newblock \showarticletitle{Social Networks and Status Attainment}.
\newblock \bibinfo{journal}{\emph{Annual Review of Sociology}}
  \bibinfo{volume}{25}, \bibinfo{number}{1} (\bibinfo{year}{1999}),
  \bibinfo{pages}{467--487}.
\newblock


\bibitem[Menon and Pfeffer(2003)]%
        {menon2003valuing}
\bibfield{author}{\bibinfo{person}{Tanya Menon} {and} \bibinfo{person}{Jeffrey
  Pfeffer}.} \bibinfo{year}{2003}\natexlab{}.
\newblock \showarticletitle{Valuing Internal vs. External Knowledge: Explaining
  the Preference for Outsiders}.
\newblock \bibinfo{journal}{\emph{Manag. Sci.}}  \bibinfo{volume}{49}
  (\bibinfo{year}{2003}), \bibinfo{pages}{497--513}.
\newblock


\bibitem[Mills(1848)]%
        {miller1848}
\bibfield{author}{\bibinfo{person}{John~Stuart Mills}.}
  \bibinfo{year}{1848}\natexlab{}.
\newblock \bibinfo{booktitle}{\emph{Principles of Political Economy}}.
\newblock \bibinfo{publisher}{Augustus M. Kelley}, \bibinfo{address}{Fairchild,
  N.J.}
\newblock


\bibitem[Morgan et~al\mbox{.}(2018)]%
        {morgan2018}
\bibfield{author}{\bibinfo{person}{Allison~C. Morgan},
  \bibinfo{person}{Dimitrios Economou}, \bibinfo{person}{Samuel~F. Way}, {and}
  \bibinfo{person}{Aaron Clauset}.} \bibinfo{year}{2018}\natexlab{}.
\newblock \showarticletitle{Prestige drives epistemic inequality in the
  diffusion of scientific ideas}.
\newblock \bibinfo{journal}{\emph{EPJ Data Science}}  \bibinfo{volume}{7}
  (\bibinfo{year}{2018}), \bibinfo{pages}{40}.
\newblock


\bibitem[Rahmattalabi et~al\mbox{.}(2021)]%
        {rahmattalabi2020}
\bibfield{author}{\bibinfo{person}{Aida Rahmattalabi}, \bibinfo{person}{Shahin
  Jabbari}, \bibinfo{person}{Himabindu Lakkaraju}, \bibinfo{person}{Phebe
  Vayanos}, \bibinfo{person}{Max Izenberg}, \bibinfo{person}{Ryan Brown},
  \bibinfo{person}{Eric Rice}, {and} \bibinfo{person}{Milind Tambe}.}
  \bibinfo{year}{2021}\natexlab{}.
\newblock \showarticletitle{Fair Influence Maximization: {A} Welfare
  Optimization Approach}.
\newblock \bibinfo{journal}{\emph{Proceedings of the AAAI Conference on
  Artificial Intelligence}} (\bibinfo{year}{2021}).
\newblock


\bibitem[Ramachandran et~al\mbox{.}(2021)]%
        {Gage2020}
\bibfield{author}{\bibinfo{person}{Govardana~Sachithanandam Ramachandran},
  \bibinfo{person}{Ivan Brugere}, \bibinfo{person}{Lav~R. Varshney}, {and}
  \bibinfo{person}{Caiming Xiong}.} \bibinfo{year}{2021}\natexlab{}.
\newblock \showarticletitle{GAEA: Graph Augmentation for Equitable Access via
  Reinforcement Learning}. In \bibinfo{booktitle}{\emph{Proceedings of the 2021
  AAAI/ACM Conference on AI, Ethics, and Society}}.
  \bibinfo{publisher}{Association for Computing Machinery},
  \bibinfo{address}{New York, NY, USA}, \bibinfo{pages}{884–894}.
\newblock


\bibitem[Rawls(2009)]%
        {rawls2009theory}
\bibfield{author}{\bibinfo{person}{J. Rawls}.} \bibinfo{year}{2009}\natexlab{}.
\newblock \bibinfo{booktitle}{\emph{A Theory of Justice}}.
\newblock \bibinfo{publisher}{Harvard University Press}.
\newblock
\showISBNx{9780674042582}


\bibitem[Rogers(1962)]%
        {rogers1962diffusion}
\bibfield{author}{\bibinfo{person}{E.M. Rogers}.}
  \bibinfo{year}{1962}\natexlab{}.
\newblock \bibinfo{booktitle}{\emph{Diffusion of Innovations}}.
\newblock \bibinfo{publisher}{Free Press of Glencoe}.
\newblock
\showISBNx{9780598411044}
\showLCCN{62015348}


\bibitem[SA and RA(2019)]%
        {CDCmanual}
\bibfield{author}{\bibinfo{person}{Rasmussen SA} {and} \bibinfo{person}{Goodman
  RA}.} \bibinfo{year}{2019}\natexlab{}.
\newblock \bibinfo{booktitle}{\emph{The CDC Field Epidemiology Manual}}.
\newblock \bibinfo{publisher}{New York: Oxford University Press}.
\newblock


\bibitem[Schelling(1971)]%
        {schelling1971}
\bibfield{author}{\bibinfo{person}{Thomas~C. Schelling}.}
  \bibinfo{year}{1971}\natexlab{}.
\newblock \showarticletitle{Dynamic models of segregation}.
\newblock \bibinfo{journal}{\emph{The Journal of Mathematical Sociology}}
  \bibinfo{volume}{1}, \bibinfo{number}{2} (\bibinfo{year}{1971}),
  \bibinfo{pages}{143--186}.
\newblock


\bibitem[Stoica and Chaintreau(2019)]%
        {stoica2019fairness}
\bibfield{author}{\bibinfo{person}{Ana-Andreea Stoica} {and}
  \bibinfo{person}{Augustin Chaintreau}.} \bibinfo{year}{2019}\natexlab{}.
\newblock \showarticletitle{Fairness in Social Influence Maximization}. In
  \bibinfo{booktitle}{\emph{Companion Proceedings of The 2019 World Wide Web
  Conference}}. \bibinfo{publisher}{Association for Computing Machinery},
  \bibinfo{pages}{569–574}.
\newblock
\showISBNx{9781450366755}


\bibitem[Stoica and Riederer(2018)]%
        {stoicaGlass}
\bibfield{author}{\bibinfo{person}{Ana-Andreea Stoica} {and}
  \bibinfo{person}{Christopher~J. Riederer}.} \bibinfo{year}{2018}\natexlab{}.
\newblock \showarticletitle{Algorithmic Glass Ceiling in Social Networks: The
  effects of social recommendations on network diversity}.
\newblock \bibinfo{journal}{\emph{WWW '18: Proceedings of the 2018 World Wide
  Web Conference}}, \bibinfo{pages}{923--932}.
\newblock


\bibitem[Sugimoto(2021)]%
        {sugimoto2021science}
\bibfield{author}{\bibinfo{person}{Cassidy~R. Sugimoto}.}
  \bibinfo{year}{2021}\natexlab{}.
\newblock \bibinfo{title}{Scientific success by numbers}.
\newblock
\newblock
\urldef\tempurl%
\url{https://www.nature.com/articles/d41586-021-01169-7}
\showURL{%
\tempurl}


\bibitem[Tang et~al\mbox{.}(2014)]%
        {TangXS14}
\bibfield{author}{\bibinfo{person}{Youze Tang}, \bibinfo{person}{Xiaokui Xiao},
  {and} \bibinfo{person}{Yanchen Shi}.} \bibinfo{year}{2014}\natexlab{}.
\newblock \showarticletitle{Influence Maximization: Near-Optimal Time
  Complexity Meets Practical Efficiency}. In
  \bibinfo{booktitle}{\emph{Proceedings of the 2014 ACM SIGMOD International
  Conference on Management of Data}}. \bibinfo{publisher}{Association for
  Computing Machinery}, \bibinfo{pages}{75--86}.
\newblock


\bibitem[Tsang et~al\mbox{.}(2019)]%
        {tsang2019group}
\bibfield{author}{\bibinfo{person}{Alan Tsang}, \bibinfo{person}{Bryan Wilder},
  \bibinfo{person}{Eric Rice}, \bibinfo{person}{Milind Tambe}, {and}
  \bibinfo{person}{Yair Zick}.} \bibinfo{year}{2019}\natexlab{}.
\newblock \showarticletitle{Group-fairness in influence maximization}. In
  \bibinfo{booktitle}{\emph{Proc. of the Int'l Joint Conf. on Artificial
  Intelligence}}. AAAI Press, \bibinfo{pages}{5997--6005}.
\newblock


\bibitem[Ugander et~al\mbox{.}(2011)]%
        {Ugander2011TheAO}
\bibfield{author}{\bibinfo{person}{Johan Ugander}, \bibinfo{person}{Brian
  Karrer}, \bibinfo{person}{Lars Backstrom}, {and} \bibinfo{person}{Cameron
  Marlow}.} \bibinfo{year}{2011}\natexlab{}.
\newblock \showarticletitle{The Anatomy of the Facebook Social Graph}.
\newblock \bibinfo{journal}{\emph{arXiv}}  \bibinfo{volume}{abs/1111.4503}
  (\bibinfo{date}{11} \bibinfo{year}{2011}).
\newblock


\bibitem[Valente and Pumpuang(2007)]%
        {identifyingValente2007}
\bibfield{author}{\bibinfo{person}{Thomas~W. Valente} {and}
  \bibinfo{person}{Patchareeya Pumpuang}.} \bibinfo{year}{2007}\natexlab{}.
\newblock \showarticletitle{Identifying Opinion Leaders to Promote Behavior
  Change}.
\newblock \bibinfo{journal}{\emph{Health Education \& Behavior}}
  \bibinfo{volume}{34}, \bibinfo{number}{6} (\bibinfo{year}{2007}),
  \bibinfo{pages}{881--896}.
\newblock


\bibitem[von Hippel(1988)]%
        {von1988sources}
\bibfield{author}{\bibinfo{person}{E. von Hippel}.}
  \bibinfo{year}{1988}\natexlab{}.
\newblock \bibinfo{booktitle}{\emph{The Sources of Innovation}}.
\newblock \bibinfo{publisher}{Oxford University Press}.
\newblock
\showISBNx{9780195094220}


\bibitem[Wang et~al\mbox{.}(2021)]%
        {Wang2021Information}
\bibfield{author}{\bibinfo{person}{Xindi Wang}, \bibinfo{person}{Onur Varol},
  {and} \bibinfo{person}{Tina Eliassi-Rad}.} \bibinfo{year}{2021}\natexlab{}.
\newblock \showarticletitle{Information Access Equality on Network Generative
  Models}.
\newblock \bibinfo{journal}{\emph{ArXiv}}  \bibinfo{volume}{abs/2107.02263}
  (\bibinfo{year}{2021}).
\newblock


\bibitem[Way et~al\mbox{.}(2016)]%
        {way2016}
\bibfield{author}{\bibinfo{person}{Samuel~F. Way}, \bibinfo{person}{Daniel~B.
  Larremore}, {and} \bibinfo{person}{Aaron Clauset}.}
  \bibinfo{year}{2016}\natexlab{}.
\newblock \showarticletitle{Gender, Productivity, and Prestige in Computer
  Science Faculty Hiring Networks}. In \bibinfo{booktitle}{\emph{Proceedings of
  the 25th International Conference on World Wide Web}}.
  \bibinfo{publisher}{International World Wide Web Conferences},
  \bibinfo{pages}{1169–1179}.
\newblock
\showISBNx{9781450341431}


\bibitem[Wilder et~al\mbox{.}(2020)]%
        {clinicalWiler2020}
\bibfield{author}{\bibinfo{person}{Bryan Wilder}, \bibinfo{person}{Laura
  Onasch-Vera}, \bibinfo{person}{Graham Diguiseppi}, \bibinfo{person}{Robin
  Petering}, \bibinfo{person}{Chyna Hill}, \bibinfo{person}{Amulya Yadav},
  \bibinfo{person}{Eric Rice}, {and} \bibinfo{person}{Milind Tambe}.}
  \bibinfo{year}{2020}\natexlab{}.
\newblock \showarticletitle{Clinical trial of an AI-augmented intervention for
  HIV prevention in youth experiencing homelessness}.
\newblock \bibinfo{journal}{\emph{arXiv}}  \bibinfo{volume}{abs/2009.09559}
  (\bibinfo{year}{2020}).
\newblock


\bibitem[Wilder et~al\mbox{.}(2018)]%
        {endtoendWilder2018}
\bibfield{author}{\bibinfo{person}{Bryan Wilder}, \bibinfo{person}{Laura
  Onasch-Vera}, \bibinfo{person}{Juliana Hudson}, \bibinfo{person}{Jose Luna},
  \bibinfo{person}{Nicole Wilson}, \bibinfo{person}{Robin Petering},
  \bibinfo{person}{Darlene Woo}, \bibinfo{person}{Milind Tambe}, {and}
  \bibinfo{person}{Eric Rice}.} \bibinfo{year}{2018}\natexlab{}.
\newblock \showarticletitle{End-to-End Influence Maximization in the Field}. In
  \bibinfo{booktitle}{\emph{International Conference on Autonomous Agents and
  Multiagent Systems (AAMAS-18)}}.
\newblock


\bibitem[Yadav et~al\mbox{.}(2018)]%
        {bridgingYadav2018}
\bibfield{author}{\bibinfo{person}{Amulya Yadav}, \bibinfo{person}{Bryan
  Wilder}, \bibinfo{person}{Eric Rice}, \bibinfo{person}{Robin Petering},
  \bibinfo{person}{Jaih Craddock}, \bibinfo{person}{Amanda Yoshioka-Maxwell},
  \bibinfo{person}{Mary Hemler}, \bibinfo{person}{Laura Onasch-Vera},
  \bibinfo{person}{Milind Tambe}, {and} \bibinfo{person}{Darlene Woo}.}
  \bibinfo{year}{2018}\natexlab{}.
\newblock \showarticletitle{Bridging the gap between theory and practice in
  influence maximization: Raising awareness about HIV among homeless youth}. In
  \bibinfo{booktitle}{\emph{Proceedings of the Twenty-Seventh International
  Joint Conference on Artificial Intelligence, {IJCAI-18}}}.
  \bibinfo{publisher}{International Joint Conferences on Artificial
  Intelligence}, \bibinfo{pages}{5399--5403}.
\newblock


\end{thebibliography}

\clearpage

%% If your work has an appendix, this is the place to put it.
\appendix

\end{document}